\documentclass[lettersize,journal]{IEEEtran}
\usepackage{amsmath,amsfonts}
\usepackage{algorithmic}
\usepackage{array}
\usepackage[caption=false,font=footnotesize,labelfont=rm,textfont=rm]{subfig}
\usepackage{textcomp}
\usepackage{stfloats}
\usepackage{url}
\usepackage{verbatim}
\usepackage{graphicx}
\usepackage{amsmath}
\usepackage{bm}
\usepackage{amssymb}
\usepackage{algorithmic}
\usepackage{algorithm}
\usepackage{amsthm}
\usepackage{color}
\allowdisplaybreaks[4]

\def\BibTeX{{\rm B\kern-.05em{\sc i\kern-.025em b}\kern-.08em
    T\kern-.1667em\lower.7ex\hbox{E}\kern-.125emX}}
\usepackage{balance}
\begin{document}
\title{Reconfigurable Intelligent Surface-Assisted Localization in OFDM Systems with Carrier Frequency Offset and Phase Noise}

\author{Hanfu Zhang, Erwu Liu, \emph{Senior Member, IEEE}, Rui Wang, \emph{Senior Member, IEEE}, Wei Ni, \emph{Fellow, IEEE}, \\Zhe Xing, Yan Liu, \emph{Member, IEEE}, and Abbas Jamalipour, \emph{Fellow, IEEE}
\thanks{
\copyright~2025 IEEE. Personal use of this material is permitted. Permission from IEEE must be obtained for all other uses, in any current or future media, including reprinting/republishing this material for advertising or promotional purposes, creating new collective works, for resale or redistribution to servers or lists, or reuse of any copyrighted component of this work in other works.

H. Zhang, E. Liu, R. Wang, and Y. Liu are with the College of Electronic and Information Engineering, Tongji University, Shanghai 201804, China. 
R. Wang is also with the Shanghai Institute of Intelligent Science and Technology, Tongji University, Shanghai 201804, China.
(e-mail: hanfuzhang@tongji.edu.cn; erwu.liu@ieee.org; ruiwang@tongji.edu.cn; yanliu2022@tongji.edu.cn).

W. Ni is with Data61, Commonwealth Science and Industrial Research Organisation, Marsfield, NSW 2122, Australia, and the School of Computer Science and Engineering, University of New South Wales, Kensington, NSW 2052, Australia (email: wei.ni@ieee.org).

Z. Xing is with the School of Information Science and Technology, North China University of Technology, Beijing 100144, China (e-mail: zxing@ncut.edu.cn).

A. Jamalipour is with the School of Electrical and Computer Engineering, University of Sydney, Sydney, NSW 2006, Australia (email: a.jamalipour@ieee.org).

}}

\maketitle
\begin{abstract}

\emph{Reconfigurable intelligent surface} (RIS)-assisted communication systems have been extensively studied for providing high-precision location services. However, most studies have overlooked the impact of \emph{carrier frequency offset} (CFO) and \emph{phase noise} (PN) resulting from hardware impairments on localization. This paper presents a novel, {\emph{alternating optimization} (AO)-based} algorithm to jointly estimate the CFO, PN, and {\color{black}\emph{user equipment} (UE) position in \emph{orthogonal frequency division multiplexing} (OFDM) systems}, where, provided the {\color{black}UE} position, closed-form expressions for the CFO and PN are derived per iteration, significantly reducing the complexity and {enhancing the stability} of the algorithm. Another important aspect is a new RIS phase shift optimization algorithm developed to minimize the analytical lower bound of localization accuracy, hence benefiting localization. The semidefinite relaxation method and Schur complement are utilized to convexify this challenging non-convex optimization problem to a semidefinite program. Simulations demonstrate the effectiveness of the proposed algorithms, with the localization accuracy enhanced by two orders of magnitude. The localization accuracy of the proposed algorithm is close to the analytical lower bound, with a root mean square error of lower than $\rm 10^{-2} \: m$.

\end{abstract}

\begin{IEEEkeywords}
Localization, reconfigurable intelligent surface, phase shift {configuration}, carrier frequency offset, phase noise.
\end{IEEEkeywords}

\section{Introduction}



\subsection{Motivation and Challenges}

\IEEEPARstart{W}{ith} the development of {\color{black}the} industrial Internet of Things, there is an increasing demand for centimeter-level localization accuracy in the \emph{sixth-generation} (6G) mobile networks \cite{letaief2019roadmap}. While \emph{global navigation satellite systems} (GNSSs) have been widely {applied to} outdoor scenarios, they undergo insufficient localization accuracy and signal instability arising from obstacles \cite{Yulong2022TASE, montenbruck2017multi}. {\color{black}Targeting at areas with little access to satellite signals, wireless localization \cite{Shaoshuai2022TCOM}, also known as communication-assisted localization~\cite{liu2022integrated}, has been widely studied. It can be potentially an integral part of \emph{integrated communication and sensing} (ISAC) systems \cite{Boyang2024TCOM,Savkin2023TVT} by utilizing communication devices and signals
to achieve localization functionality.}

{\color{black}While wireless localization systems} have been extensively studied, few studies have considered the {adverse effects} of hardware impairments, such as \emph{carrier frequency offset} (CFO) and \emph{phase noise} (PN), on localization accuracy. CFO is a constant parameter {due to} the frequency {offset} between the transmitter and the receiver. {Doppler shift can also lead to CFO}. PN is a random process {resulting from} the fluctuation of the {oscillators at the transmitters and receivers} \cite{wu2004ofdm}. CFO and PN can result in signal constellation rotations in single carrier communication systems, or \emph{inter-carrier interference} (ICI) in {broadband communication systems, for example,} \emph{orthogonal frequency division multiplexing} (OFDM) systems. The overall communication performance and the localization accuracy can degrade significantly, due to the CFO and PN.

Another challenge arises from the increasingly likely blockages of signals, e.g.,  millimeter wave (mmWave) signals with the quasi-optical property, in many dense urban areas~\cite{JinXin2024TWC}. \emph{Reconfigurable intelligent surface} (RIS) 
has been considered in communication systems to deal with the problem of signal blockage~\cite{Xin2023TIFS, Jingheng2022TWC}. Localization can be implemented with the help of virtual \emph{line-of-sight} (LoS) paths provided by RISs even when the LoS paths are blocked, and can be improved via passive beamforming. In this sense, RIS-assisted {\color{black} \emph{user equipment} (UE)} localization is of practical interest, and could be particularly challenging with the further consideration of the CFO and PN. However, this has not been studied in the existing literature.

\subsection{Related works}

\subsubsection{\color{black}RIS-free localization with ideal hardware}
Communication systems have been widely employed to assist in high-precision {\color{black} UE} localization, typically under a far-field setting; i.e., the {\color{black} UE} and the anchors are in the far field of each other. For example, Shahmansoori \emph{et al.} \cite{shahmansoori2017position} proposed a three-stage localization algorithm that improves the simultaneous orthogonal matching pursuit algorithm and the space alternating generalized expectation-maximization algorithm, and exploited the sparsity of millimeter-wave \emph{multiple-input multiple-output} (MIMO) channels to simplify the channel model. Hu \emph{et al.} \cite{hu2014esprit} presented a \emph{two-dimensional} (2D) localization algorithm based on the celebrated \emph{estimating signal parameters via rotational invariance technique} (ESPRIT), which estimates {\color{black} UE} positions by constructing signal subspaces. 

\subsubsection{\color{black}RIS-free localization under CFO and PN}
Some studies \cite{meles2023impact,chen2022mcrb} have explored the impact of realistic hardware impairments, including the CFO and PN, with no RIS considered.
For instance, Meles \emph{et al.} \cite{meles2023impact} simulated the effects of CFO and PN on \emph{angle-of-arrival} (AOA) estimation. {\color{black}Nevertheless, CFO and PN were ignored when designing the algorithms and were only included in simulations, showing that the CFO and PN can significantly and adversely affect localization and tracking performance.} Chen \emph{et al.}~\cite{chen2022mcrb} considered the impacts of CFO, PN, mutual coupling, and power amplifier nonlinearity on 6G localization, and derived the misspecified \emph{Cram\'{e}r-Rao lower bound} (CRLB). {\color{black}It was analytically concluded that the PN affects both angle and delay estimation, whereas the CFO has a more significant effect on angle estimation.} However, neither of {\color{black} these works, i.e., \cite{meles2023impact, chen2022mcrb}} have provided an estimation algorithm to handle hardware impairments. 

Some other works \cite{qiu2020aoa,zakariyya2023joint,xiao2023improved,zheng2020joint} have delved into localization algorithm design with either CFO or PN. 
On the one hand, Qiu \emph{et al.} \cite{qiu2020aoa} designed a \emph{maximum likelihood estimation} (MLE)-based weighted iterative algorithm for the joint estimation of CFO and AOA. Zakariyya \emph{et al.}~\cite{zakariyya2023joint} proposed a high-resolution {\color{black}enhancement of the} \emph{multiple signal classification} (MUSIC) algorithm for localization. {\color{black}While the algorithm balanced spectrum resolution and accuracy, the CFO was not estimated in localization, leading to a loss in localization accuracy.} On the other hand, Xiao \emph{et al.} \cite{xiao2023improved} combined the \emph{expectation maximization} (EM) and MUSIC algorithms for joint estimation of AOA and PN. {\color{black}At a typical \emph{signal-to-noise ratio} (SNR) level, this algorithm can reduce the estimation error by over $55\%$ compared to its benchmarks. However, only a simple single-carrier system under a 2D far-field setting was considered}. Zheng \emph{et al.} \cite{zheng2020joint} formulated a \emph{maximum a-posteriori} (MAP) problem solved iteratively using the block \emph{majorization minimization} (MM) algorithm after approximation. {\color{black}Under a medium or small PN, the algorithm improves localization and channel estimation. 
However, none of these methods \cite{qiu2020aoa,zakariyya2023joint,xiao2023improved,zheng2020joint}  have considered the use of RISs or addressed the signal blockage by obstacles.}

\subsubsection{\color{black}RIS-assisted localization with ideal hardware}
RIS has been integrated into localization systems to address the problem of signal blockage, but little consideration has been given to the above-mentioned hardware impairments, i.e., CFO and PN. 
From the perspective of localization distance, existing research can be divided between far-field localization \cite{wang2022location, elzanaty2021reconfigurable, keykhosravi2021semi} and near-field localization \cite{han2022localization, wang2022location, elzanaty2021reconfigurable, zhang2022hybrid}. In the far-field case, the {{\color{black} UE}-RIS} distance exceeds Fraunhofer distance \cite{balanis2016antenna}, and the signal can be approximated as a planar wave. Conversely, in the near-field case, the signal cannot be treated as parallel propagation and should be considered as a spherical wave.
For example, Luan \emph{et al.} \cite{luan2021phase} focused on the near-field  2D localization scenario, and proposed a robust optimization {algorithm} to {configure} the RIS phase shifts for improving localization accuracy. The transmit powers were also optimized in the considered localization system. Ozturk \emph{et al.} \cite{ozturk2023ris} introduced a more practical RIS model with phase-dependent amplitude variations in the near-field circumstance. The mis-specified CRLB was derived and an approximated mismatched maximum likelihood-based algorithm was proposed for joint UE localization and RIS parameter estimation. Lin \emph{et al.} \cite{lin2021channel} introduced a novel twin-RIS configuration comprising two RISs with relative spatial rotations. To extract the parameters of the cascaded channel, the authors employed array signal processing and atomic norm denoising techniques. They established nonlinear equations {exploiting} the characteristics of the twin-RIS structure to recover the multipath parameters. Localization for  {\color{black} UE}s in the far-field scenario was then performed based on the estimated channel parameters.

\subsubsection{\color{black}RIS-assisted localization under CFO and PN}
Most existing RIS-assisted localization studies have assumed ideal channel conditions without considering realistic hardware impairments. Only several studies have considered modeling some specific impairments. For example, the studies in~\cite{elzanaty2021reconfigurable, keykhosravi2021semi, dardari2021nlos} took into account clock offset, and the work in~\cite{dardari2021nlos} considered phase offsets. Zhang \emph{et al.} \cite{zhang2024user} considered the CFO while overlooking the effects of PN. They proposed a MUSIC-based algorithm that directly estimates the position parameters together with the CFO. However, the algorithm is only suitable for far-field scenarios. 

{\color{black}
Unlike these existing works, we not only analytically quantify the impact of CFO and PN on localization accuracy, but also design efficient localization and RIS configuration algorithms. As a result, the localization accuracy can be improved by two orders of magnitude, compared to the existing methods.} To the best of our knowledge, no existing research has specifically analyzed the impact of both CFO and PN on RIS-assisted near-field localization systems {\color{black}nor developed corresponding localization and RIS configuration algorithms}.

\subsection{Contributions}
In this paper, we {investigate} a RIS-assisted OFDM system, where practical CFO and PN convolve with the data symbols, causing signal constellation rotation and ICI and, consequently, hindering {\color{black} UE} localization in near-field environments. The key contributions of this paper are outlined as follows:


\begin{itemize}
    \item We investigate a problem of  near-field localization in a RIS-{aided} OFDM system, while {taking into account} the CFO and PN. We propose an \emph{alternating optimization} (AO)-based iterative estimation approach, where the MAP criterion and \emph{gradient descent} (GD) {algorithm} are employed 
    with closed-form expressions derived for the CFO and PN estimation.
    
    \item We derive the \emph{hybrid CRLB} (HCRLB) for the {considered joint estimation problem. The phase ambiguity in CFO and PN estimation is analyzed}. To evaluate the accuracy of joint CFO and PN estimation, a joint \emph{mean square error} (MSE) metric is introduced. 

    \item We further optimize the phase shifts of the RIS to improve {the performance of} localization. By applying the \emph{semi-definite relaxation} (SDR) and the Schur complement, we convert the non-convex optimization problem into a convex \emph{semi-definite programming} (SDP) problem solved efficiently.

    \item We {perform comprehensive} numerical simulations to assess the performance of the proposed algorithms. It is shown that the localization accuracy {is} improved by two orders of magnitude compared to random RIS phase shift configuration. Compared to the state-of-the-art benchmarks, our algorithms can improve the localization accuracy by about one order of magnitude under relatively low transmit power and even by two orders of magnitude under high transmit power.
     
\end{itemize}

\begin{figure}[!t]
\centering
\includegraphics[width=0.9\linewidth]{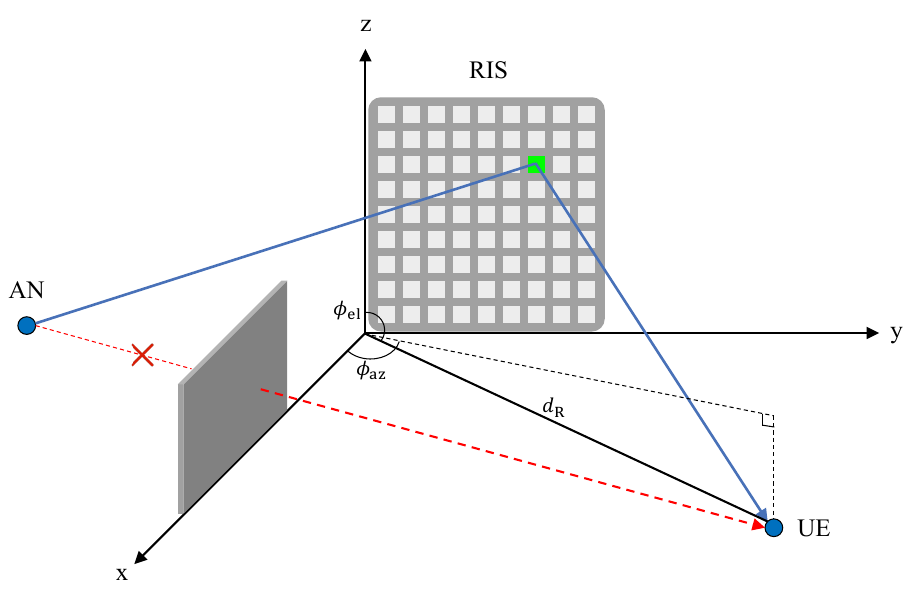}
\caption{The considered RIS-aided 3D near-field localization system, including an AN, a UE, and an ${\color{black} N_{\text{R}}}$-element RIS. The LoS link between the AN and UE is {obstructed}. The UE utilizes the signals transmitted by the AN and reflected by the RIS to localize its position.}
\label{SystemModel}
\end{figure}

The rest of this paper is {organized} as follows. Section~II {presents} the system model. Section III presents the joint estimation algorithm. Section IV derives the HCRLB for the joint estimation algorithm. The proposed RIS phase shift optimization method is delineated in Section V. Section VI showcases comprehensive simulation results, and the paper is concluded in Section VII.

\emph{Notation}: $\mathbb{E} (\cdot)$ and $p(\cdot)$ denote expectation and \emph{probability density function} (PDF), respectively. ${\rm tr} (\cdot)$, $\det(\cdot)$, $(\cdot)^\top$, $(\cdot)^*$, $(\cdot)^{\rm H}$, and $(\cdot)^{\rm -1}$ denote trace, determinant, transpose, conjugate, hermitian, and inverse, respectively. ${\rm Diag} (\bm{v})$ indicates a diagonal matrix {where the entries of vector $\bm{v}$ appear on its main diagonal}. $\bm{v}[n]$ denotes the $n$-th entry of vector $\bm{v}$. ${\rm Blkdiag} (\mathbf{M}_0, \mathbf{M}_1, \cdots, \mathbf{M}_{N-1})$ denotes a block diagonal matrix with $\mathbf{M}_0, \mathbf{M}_1, \cdots, \mathbf{M}_{N-1}$ being the matrices along its main diagonal. $\left[ \mathbf{M} \right]_{i:j, :}$ refers to a sub-matrix encompassing rows $i$ to $j$ of matrix $\mathbf{M}$, while $\left[ \mathbf{M} \right]_{:, i:j}$ denotes a sub-matrix encompassing columns $i$ to $j$ of matrix $\mathbf{M}$. $\bm{0}_{M \times N}$ and $\mathbf{I}_M$ denote the $M \times N$ all-zero matrix and the $M \times M$ identity matrix, respectively. $\Re \left\{\cdot \right \}$ and $\Im \left\{\cdot \right \}$ represent the real and imaginary components of a complex value, respectively. $\mathbb{C}^{m \times n}$ and $\mathbb{R}^{m \times n}$ collect all $m \times n$ matrices with complex and real entries, respectively. $\mathcal{CN} (\bm{\mu}, \mathbf{\Sigma})$ and $\mathcal{N} (\bm{\mu}, \mathbf{\Sigma})$ represent complex and real Gaussian distributions, respectively, with a mean vector $\bm{\mu}$ and covariance matrix $\mathbf{\Sigma}$. Unless otherwise specified, all indices in the paper start from~0. Moreover, we summarize the notation used in this paper in Table \ref{tab_notation}.

\begin{table}
\caption{Notation and definition.}
\label{tab_notation}
\centering
\begin{tabular}{| c | l |}
\hline
Notation & Definition \\
\hline
${\color{black} N_{\text{R}}}$  & Number of RIS elements \\
$N$ & Number of subcarriers \\
$L$ & Length of the approximate PN vector \\
$d$ & Element spacing of the RIS \\
$D$ & Aperture of the RIS \\
${\color{black} f_{\text{c}}}$ & Carrier frequency of the system \\
$B$ & Bandwidth of the system\\
$\lambda$ & Wavelength of the system \\
$c$ & Speed of light \\
$P$ & Transmit power at the AN \\
$\Delta f$ & CFO  \\
$\phi$ & Normalized CFO \\
$\color{black}S$ & Position sampling amount \\
$\sigma^2$ & Variance of the noise at the UE \\
$\sigma_\Delta^2$ & Innovation variance for PN \\
${\color{black} \bm{p}_{\text{a}} }$ & Coordinate of the AN \\
$\bm{p}_r$ & Coordinate of the {\color{black}$r$-th element} of the RIS \\
$\bm{p}_{\color{black}\text{u}}$  & Coordinate of the UE position \\
$\bm{\theta}$ & PN vector \\
$\bm{\eta}$  & Approximate PN vector \\
$\bm{y}$ & Received signal at the UE \\
$\bm{\mu}$ & Noise-free received signal at UE \\
$\bm{h}$  & AN-RIS-UE frequency-domain channel \\
$\bm{w}$ & RIS phase shift vector \\
$\mathbf{s}$ & Transmitted symbol sequence at the AN \\
$\bm{v}$ & Additive noise at UE \\
$\bm{\lambda}$ & Parameter vector to be estimated \\
$\bm{\alpha}$ & Transformed parameter vector \\
$\mathbf{B}$  & Bayesian information matrix \\
$\bm{\Lambda}_\phi$ & CFO matrix \\
$\bm{\Lambda}_{\bm{\theta}}$ & PN matrix \\
$\mathbf{\Pi}$ & Transition matrix for the PN vector\\
$\mathbf{\Xi}$ & Transition matrix for the HCRLB \\
\hline
\end{tabular}
\end{table}

\section{System Model}
\label{section: system_model}

The considered RIS-assisted OFDM downlink wireless localization system is composed of an \emph{anchor node} (AN), a {\color{black} UE}, and an {\color{black}$N_{\text{R}}$}-element RIS, {as illustrated in Fig.~\ref{SystemModel}}. We {make the assumption} that the LoS path between the AN and UE is {obstructed}, and the \emph{channel state information} (CSI) remains unchanged during localization, as in \cite{wang2022location, luan2021phase, dardari2021nlos}. {\color{black}Since both the AN and RIS are part of the network infrastructure, the }positions of the AN and each {RIS element}, denoted as ${\color{black} \bm{p}_{\text{a}} }$ and $\bm{p}_{r} = [x_r, y_r, z_r]^\top$, respectively, are known{\color{black}, as typically considered in RIS-assisted localization studies, e.g., \cite{elzanaty2021reconfigurable, luan2021phase, dardari2021nlos, teng2022bayesian}. Here, $r$ is the index of a RIS element}. The position of UE, denoted as $\bm{p}_{\color{black}\text{u}} = [x_{\color{black}\text{u}}, y_{\color{black}\text{u}}, z_{\color{black}\text{u}}]^\top$, is to be estimated.

The RIS elements are organized in a \emph{uniform planar array} (UPA). We take the bottom-left corner of the RIS as the reference point and place it at the coordinate origin. The RIS is placed on the $yz$-plane, and both edges on the coordinate axes are located in the positive half-axis. Hence, the coordinates of the $r$-th RIS element can be represented as {\color{black}$\bm{p}_r = [0, d \times (r \mod \sqrt{N_{\text{R}}}), d \times \lceil \frac{r+1}{\sqrt{N_{\text{R}}}} - 1 \rceil]^\top$}, where $d$ stands for the element spacing of the RIS. {\color{black} We set $d$ to half-wavelength [23], [24], [33], because signals at different RIS elements may couple if $d$ is too small, or the RIS array's sidelobe can be significant, leading to poor directionality for the RIS if $d$ is too large.} To satisfy the conditions of the near-field setup, both the {AN-RIS distance}, as well as the {RIS-UE} distance, should lie within the Fresnel region \cite{balanis2016antenna}, i.e., the interval $[ 0.62 \sqrt{D^3 / \lambda}, 2 D^2 / \lambda ]$, where $D = \sqrt{2} d (\sqrt{{\color{black} N_{\text{R}}}} - 1)$ represents the aperture of the RIS and $\lambda$ {denotes} the wavelength.

Additionally, we adopt a polar coordinate system for estimating the {\color{black} UE}'s position, i.e., using ${\color{black}d_\text{R}}$, ${\color{black}\phi_\text{az}}$ and ${\color{black}\phi_\text{el}}$, to express the {\color{black} UE}'s position, indicating the distance, azimuth angle, and elevation angle of UE relative to the coordinate origin, respectively. The relationships between these three parameters and $x_{\color{black}\text{u}}$, $y_{\color{black}\text{u}}$ and $z_{\color{black}\text{u}}$ are given by
\begin{subequations} \label{eq:cartesian_2_polar}
\begin{align}
{\color{black}d_\text{R}} &= \sqrt{x_{\color{black}\text{u}}^2 + y_{\color{black}\text{u}}^2 + z_{\color{black}\text{u}}^2}; \\
{\color{black}\phi_\text{az}} &= \arctan \frac{y_{\color{black}\text{u}}}{x_{\color{black}\text{u}}}; \\
{\color{black}\phi_\text{el}} &= \arccos \frac{z_{\color{black}\text{u}}}{\sqrt{x_{\color{black}\text{u}}^2 + y_{\color{black}\text{u}}^2 + z_{\color{black}\text{u}}^2}},
\end{align}
\end{subequations}
and, in turn, $x_{\color{black}\text{u}}$, $y_{\color{black}\text{u}}$ and $z_{\color{black}\text{u}}$ can be written as
\begin{subequations} \label{eq:polar_2_cartesian}
\begin{align}
x_{\color{black}\text{u}} &= {\color{black}d_\text{R}} \sin {\color{black}\phi_\text{el}} \cos {\color{black}\phi_\text{az}}; \\
y_{\color{black}\text{u}} &= {\color{black}d_\text{R}} \sin {\color{black}\phi_\text{el}} \sin {\color{black}\phi_\text{az}}; \\
z_{\color{black}\text{u}} &= {\color{black}d_\text{R}} \cos {\color{black}\phi_\text{el}}.
\end{align}
\end{subequations}

{An OFDM symbol with a center carrier frequency of ${\color{black} f_{\text{c}}}$, a signal bandwidth of $B$ and $N$ subcarriers} is considered in this paper. Upon sampling at a rate of ${\color{black}T_\text{s}} = \frac{1}{NB}$ and removing the \emph{cyclic prefix} (CP) at the UE, the received time-domain signal {$\bm{y}$} can be written as
\begin{align}
\label{receivedSignal}
    \bm{y} = \sqrt{P} \bm{\Lambda}_\phi \bm{\Lambda}_{\bm{\theta}} \mathbf{S} ^ \top {\color{black} \bm{h}_\text{LoS}} + \sqrt{P} \bm{\Lambda}_\phi \bm{\Lambda}_{\bm{\theta}} \mathbf{S} ^ \top {\color{black} \bm{h}_\text{NLoS}} + \tilde{\bm{v}},
\end{align}
where
\begin{itemize}
\item $P$ is the transmit power;
\item $\bm{\Lambda}_\phi$ is the CFO matrix {resulting from a frequency mismatch between the transmitter and receiver. $\bm{\Lambda}_\phi$} is a diagonal matrix with $\left[ \bm{\Lambda}_{\phi} \right]_{n,n} = e^{j 2 \pi\frac{n \Delta f }{N B}} = e^{j 2 \pi \frac{n}{N} \phi}$ being the $(n, n)$-th element along its main diagonal, where $\Delta f$ is the CFO, and $\phi = \frac{\Delta f }{B}$ is called the normalized CFO;
\item $\bm{\Lambda}_{\bm{\theta}} \triangleq {\rm Diag} \left[ e^{j \bm{\theta}[0]}, e^{j \bm{\theta}[1]}, \cdots, e^{j \bm{\theta}[N-1]} \right]$ is the PN matrix. Drawing upon the attributes of the PN observed in practical oscillators, the PN vector $\bm{\theta}$ can be modeled as a Wiener process \cite{lin2006joint}, i.e.,
\begin{equation}
 \bm{\theta}[n] = \bm{\theta}[n-1] + \bm{\Delta}[n],
\end{equation}
where $\bm{\Delta}[n]$ is a Gaussian variable, i.e., $\bm{\Delta}[n] \sim \mathcal{N}(0, \sigma_\Delta^2)$. As in \cite{lin2007variational}, we assume that $\bm{\theta}[-1] = 0$. Then, the PN vector $\bm{\theta} \triangleq \left[ \bm{\theta}[0], \bm{\theta}[1], \cdots, \bm{\theta}[N-1] \right]^\top$ yields $\bm{\theta} \sim \mathcal{N}(\bm{0}_{N \times 1}, \bm{\Psi})$, where $\mathbf{\Psi}$ is given by
\begin{equation}
    \mathbf{\Psi} = \sigma_\Delta^2
    \begin{bmatrix}
        1 & 1 & 1 & \cdots & 1 & 1 \\
        1 & 2 & 2 & \cdots & 2 & 2 \\
        1 & 2 & 3 & 3 & \cdots & 3 \\
        \vdots & \vdots & \vdots & \vdots & \ddots & \vdots \\
        1 & 2 & 3 & \cdots & N-1 & N
    \end{bmatrix};
\end{equation}

\item $\mathbf{S} \triangleq \sqrt{N} \mathbf{F}^{\rm H} {\rm Diag} (\bm{s})$, where $\mathbf{F}^{\rm H}$ is the inverse \emph{fast Fourier transform} (FFT) matrix and $\bm{s}$ is the frequency-domain transmitted symbol sequence at the AN;
\item ${\color{black} \bm{h}_\text{LoS}}$ and ${\color{black} \bm{h}_\text{NLoS}}$ represent the LoS and non-LoS parts of the frequency-domain channel, respectively {\color{black}\cite{dardari2021nlos, wymeersch2022radio}};
\item $\tilde{\bm{v}}$ is the additive Gaussian noise at the UE, and follows the distribution $\tilde{\bm{v}} \sim \mathcal{CN}(\bm{0}_{N \times 1}, {\color{black} \sigma_\text{AWGN}^2} \mathbf{I}_{N})$.
\end{itemize}

Assume that ${\color{black} \bm{h}_\text{NLoS}}$ follows a known Gaussian distribution {\color{black}\cite{hu2020location}, i.e.,} ${\color{black} \bm{h}_\text{NLoS}} \sim \mathcal{CN}({\color{black}\bm{0}_{N \times 1}}, {\color{black} \sigma_\text{NLoS}^2} \mathbf{I}_N)$. Thus, ${\color{black} \bm{\mu}_{\text{NLoS}}} \triangleq \mathbf{X} {\color{black} \bm{h}_\text{NLoS}}$ also follows a known Gaussian distribution ${\color{black} \bm{\mu}_{\text{NLoS}}} \sim \mathcal{CN}({\color{black}\bm{0}_{N \times 1}}, {\color{black} \sigma_\text{NLoS}^2} \mathbf{X} \mathbf{X}^{\rm H})$ with $\mathbf{X} \triangleq \sqrt{P} \bm{\Lambda}_\phi \bm{\Lambda}_{\bm{\theta}} \mathbf{S} ^ \top$ {\color{black}and $\mathbf{X} \mathbf{X}^{\rm H} = N P \mathbf{I}_N$}. Combine ${\color{black} \bm{\mu}_{\text{NLoS}}}$ with the additive Gaussian noise $\tilde{\bm{v}}$. {\color{black} For conciseness, we rewrite ${\color{black} \bm{h}_\text{LoS}}$ with $\bm{h}$ in the rest of this paper.} The received signal is rewritten as
\begin{equation}
    \label{receivedSignalWithoutHNLoS}
    \bm{y} = \sqrt{P} \bm{\Lambda}_\phi \bm{\Lambda}_{\bm{\theta}} \mathbf{S} ^ \top \bm{h} + \bm{v},
\end{equation}
where $\bm{v} \triangleq {\color{black} \bm{\mu}_{\text{NLoS}}} + \tilde{\bm{v}}$ {\color{black} yields a Gaussian distribution}, i.e, $\bm{v} \sim \mathcal{CN}({\color{black}\bm{0}_{N \times 1}}, \sigma^2 \mathbf{I}_N)$, with $\sigma^2 = {\color{black} \sigma_\text{AWGN}^2} + {\color{black} \sigma_\text{NLoS}^2 N P}$.

Each element of $\bm{h}$ can be written as {\color{black}\cite{luan2021phase}}
{\color{black}\begin{equation}
    \label{h_los}
    \bm{h}[n] = \sum_{r=0}^{N_{\text{R}}-1} \sqrt{\rho^{n,r}_{\text{aR}}} \sqrt{\rho_{{\text{Ru}}}^{n,r}} \bm{w}[r] e^{-j 2 \pi \frac{n B}{N} (\tau_{\text{a}r} + {\color{black} \tau_{r\text{u}}}) },
\end{equation}}where ${\color{black}\sqrt{\rho_{\text{aR}}^{n,r}}} = \frac{\sqrt{G_{\color{black}\text{a}}} \lambda_n}{4 \pi d_{{\color{black}\text{a}}r}} $ and ${\color{black}\sqrt{\rho_{\text{Ru}}^{n,r}}} = \frac{\sqrt{G_{\color{black}\text{t}}} \lambda_n}{4 \pi {\color{black} d_{r\text{u}}}} $ are the free-space path losses from the AN to the $r$-th {RIS element}, and from the $r$-th {RIS element} to the UE, respectively; $\bm{w}[r]$ is the phase shift of the $r$-th {RIS element}; $\tau_{{\color{black}\text{a}}r} = \frac{d_{{\color{black}\text{a}}r}}{c} = \frac{\Vert \bm{p}_r - {\color{black} \bm{p}_{\text{a}} } \Vert}{c}$ is the delay from the AN to the $r$-th {RIS element}, while ${\color{black} \tau_{r\text{u}}} = \frac{{\color{black} d_{r\text{u}}}}{c} = \frac{\Vert \bm{p}_{\color{black}\text{u}} - \bm{p}_r \Vert}{c}$ is the delay from the $r$-th {RIS element} to the UE; $c$ is the speed of light.

\section{Proposed Joint CFO, PN and UE Position Estimation Algorithm}
\label{section: method}
In previous works \cite{luan2021phase, dardari2021nlos}, localization was based on frequency-domain signals, and different subcarriers' signals were often assumed to be independent. However, the CFO and PN {\color{black} can result in ICI, which increases the difficulty in the estimation of the CFO, PN, and UE position.} The estimation can only be conducted in the time domain. In this section, we propose an AO-based iterative algorithm, which employs the MAP criterion and GD algorithm to jointly estimate the normalized CFO, PN, and the {UE position}. {\color{black}To address the new challenge of estimating the CFO and PN, the PN vector is approximated by exploiting the redundancy of its covariance matrix, and the posterior functions concerning the CFO and PN are linearized. Closed-form estimates are derived for both CFO and PN.}

{{\color{black}The proposed localization algorithm can serve as a prerequisite for tracking. Since the Doppler shift can be interpreted as part of CFO, by estimating the CFO, we can provide an estimate of the UE's speed and improve tracking. For example, the UE position and CFO estimates provided by our proposed localization algorithm can be used as initialization inputs for existing tracking algorithms, e.g.,~\cite{dardari2021nlos}.}

Define the parameter vector $\bm{\lambda} \triangleq \left[ \phi, \bm{\theta}^\top, {\color{black}d_\text{R}}, {\color{black}\phi_\text{az}}, {\color{black}\phi_\text{el}} \right]^\top$, wherein $\bm{\theta}$ is a random variable and the others are deterministic. {\color{black}The $k$-th variable in $\bm{\lambda}$ is denoted as $\lambda_k$.} The posterior distribution of $\bm{\lambda}$, given the received signal $\bm{y}$, can be expressed as
\begin{equation}
    \label{posterior}
    p(\bm{\lambda} \mid \bm{y}) = \frac{p(\bm{y} \mid \bm{\lambda}) \times p(\bm{\theta})}{p(\bm{y})}.
\end{equation}

According to the MAP criterion, $\bm{\lambda}$ can be estimated by optimizing the following unconstrained function:
{\color{black}\begin{equation}
    \label{objective}
    \hat{\bm{\lambda}} \propto \arg \min_{\bm{\lambda}} \: -\log p(\bm{y} \mid \bm{\lambda}) - \log p(\bm{\theta}),
\end{equation}
which is obtained by calculating the negative logarithmic function of (\ref{posterior}) and then suppressing constants.}

\subsection{Phase Noise Estimation}

\subsubsection{Approximation of PN vector}

\begin{figure}[!t]
\centering
\includegraphics[width=0.9\linewidth]{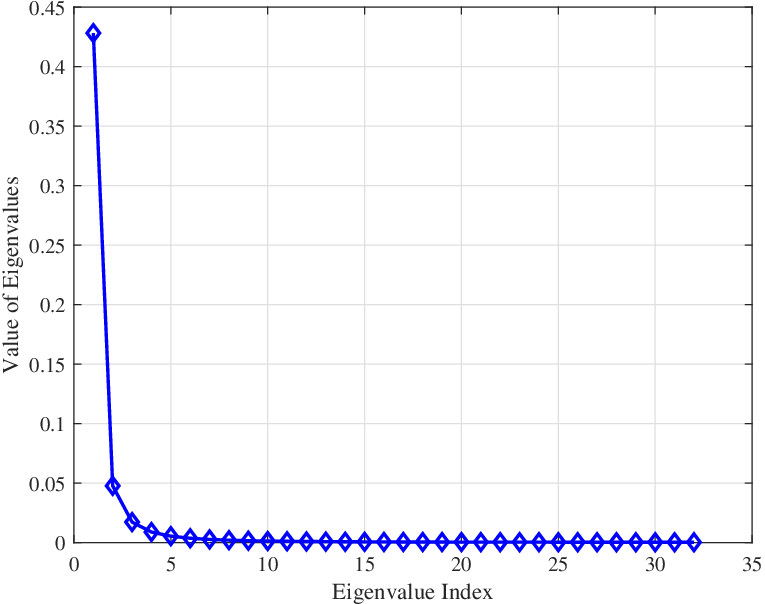}
\caption{Eigenvalues of $\bm{\Psi}$ at the number of subcarriers $N=32$, and the innovation variance for PN $\sigma_\Delta^2=10^{-3}$.}
\label{eig}
\end{figure}

From the definition of $\bm{\lambda}$, many variables need to be jointly estimated at the UE. To {lower} the computational complexity of the estimation, we exploit the redundancy of the covariance matrix of the PN vector to approximate it and reduce variables {requiring estimation}. {\color{black}This approximation is based on the commonly used truncated \emph{singular value decomposition} (SVD) method \cite{10476639, 7067415}.}

As {illustrated} in Fig. \ref{eig}, the eigenvalues of $\bm{\Psi}$ are mostly close to zero, indicating significant redundancy in $\bm{\Psi}$. With this property, we can approximate the PN vector as
\begin{equation}
\label{eta_2_theta}
    \bm{\theta} \approx \bm{\Pi} \bm{\eta},
\end{equation}
where $\bm{\eta} \sim \mathcal{N}(\bm{0}_{L \times 1}, \mathbf{I}_{L})$ contains $L \leq N$ PN variables, and $\bm{\Pi} \in \mathbb{R}^{N \times L} $ is the transition matrix. To derive $\bm{\Pi}$, we first perform \emph{eigenvalue decomposition} (EVD) on $\bm{\Psi}$. {\color{black}For the real symmetric matrix $\mathbf{\Psi}$, the EVD is equivalent to its SVD. Based on the EVD, we have} $\bm{\Psi} = \mathbf{U} \mathbf{D} \mathbf{U}^\top$, where $\mathbf{D} = {\rm Diag}(\bm{e})$ and $\bm{e}$ collects the eigenvalues of $\bm{\Psi}$ arranged in descending order. Subsequently, $\bm{\Pi}$ can be calculated by {\color{black}$\bm{\Pi} = [\mathbf{U}]_{:, 0:L-1} \times {\rm Diag} \left\{ \tilde{\bm{e}}\right\}$, with $\tilde{\bm{e}} \triangleq \left[ \sqrt{\bm{e}[0]}, \sqrt{\bm{e}[1]}, \cdots, \sqrt{\bm{e}[L-1]} \right]^\top$}.

{\color{black} Define $\tilde{\mathbf{\Psi}}=\mathbf{\Pi} \mathbf{\Pi}^\top$. Thus, $\mathbf{\Pi} \bm{\eta} \sim \mathcal{N} (\bm{0}_{N \times 1}, \tilde{\mathbf{\Psi}})$. $\tilde{\mathbf{\Psi}}$ can be rewritten as
\begin{align}
     \tilde{\mathbf{\Psi}} &= \left[\mathbf{U}\right]_{:, 0: L-1} {\rm Diag} \left\{\tilde{\bm{e}}\right\} \left\{\tilde{\bm{e}}\right\}^\top \left[\mathbf{U}\right]_{:, 0: L-1}^\top \notag \\
     &= \sum_{l=0}^{L-1} \bm{e}[l] \left[\mathbf{U}\right]_{:, l} \left[\mathbf{U}\right]_{:, l}^\top,
\end{align}
and is expected to approach $\mathbf{\Psi}$ when $L$ is sufficiently large since most of the eigenvalues of $\mathbf{\Psi}$ are close to zero. In this paper, we set $L=\frac{N}{2}$. In the following analysis, $\bm{\eta}$ is estimated, instead of $\bm{\theta}$.}

\subsubsection{Estimation of approximate PN vector}
By approximating the PN vector $\bm{\theta}$, the objective function on the right-hand side (RHS) of (\ref{objective}) can be rewritten as
\begin{align}
\label{estimation_objective_function}
    \mathcal{L}(\bm{\lambda}) &= -\log p(\bm{y} \mid \bm{\lambda}) - \log p(\bm{\eta}) \notag \\
    &\propto \frac{1}{\sigma^2} \Vert \bm{y} - \bm{\mu} \Vert ^2 + \frac{1}{2} \bm{\eta}^\top \bm{\eta},
\end{align}
where $\bm{\mu} \triangleq \sqrt{P} \bm{\Lambda}_\phi \bm{\Lambda}_{\bm{\theta}} \mathbf{S} ^ \top \bm{h} $ stands for the noise-free received signal at the UE.

{Owing to} the typically small PN variance of practical oscillators, the elements of the PN matrix $\bm{\Lambda}_{\bm{\theta}}$ can be approximated by employing its first-order Taylor expansion, resulting in $e^{j \bm{\theta}[n]} \approx 1 + j \bm{\theta}[n]$. Consequently, we can approximate $\bm{\Lambda}_{\bm{\theta}}$ as $\bm{\Lambda}_{\bm{\theta}} \approx \mathbf{I}_N + j {\rm Diag}(\bm{\Pi \eta})$. With this approximation, the objective function in (\ref{estimation_objective_function}) can be rewritten as
\begin{align}
\label{L_theta}
    \mathcal{L}_{\bm{\eta}}^{[q+1]} &\approx \frac{1}{\sigma^2} (\bar{\bm{y}} - \mathbf{Q}\bm{\eta})^{\rm H} (\bar{\bm{y}} - \mathbf{Q}\bm{\eta}) + \frac{1}{2} \bm{\eta}^\top \bm{\eta} \notag \\
    &= \frac{1}{\sigma^2} \left( 
        \bar{\bm{y}}^{\rm H} \bar{\bm{y}} 
	- 2 \Re \left\{\bar{\bm{y}}^{\rm H} \mathbf{Q}\right\} \bm{\eta} + \bm{\eta}^\top \Re \left\{\mathbf{Q}^{\rm H} \mathbf{Q} \right\} \bm{\eta}
	\right) \notag \\
    & \quad + \frac{1}{2} \bm{\eta}^\top \bm{\eta},
\end{align}
where $\bar{\bm{y}} \triangleq \bm{y} - \sqrt{P} \hat{\bm{\Lambda}}_\phi^{[q]} \mathbf{S}^\top \hat{\bm{h}}^{[q]}$, and $\mathbf{Q} \triangleq j \sqrt{P} {\rm Diag}(\hat{\bm{\Lambda}}_\phi^{[q]} \mathbf{S}^\top  \hat{\bm{h}}^{[q]} ) \bm{\Pi}$. The superscript $``^q"$ indicates the $q$-th iteration.

Setting the partial derivative of (\ref{L_theta}) {regarding} $\bm{\eta}$ to zero, 
\begin{align}
    \frac{\partial \mathcal{L}^{[q+1]}_{\bm{\eta}}}{\partial \bm{\eta}} = - \frac{2}{\sigma^2} \Re \left\{\mathbf{Q}^{\rm H} \bar{\bm{y}}  \right\} 
    + \frac{2}{\sigma^2} \Re \left\{\mathbf{Q}^{\rm H} \mathbf{Q} \right\} \bm{\eta} + \bm{\eta}
    = \bm{0}_{L \times 1},
\end{align}
we can solve
\begin{equation}
    \label{eta_es}
    \hat{\bm{\eta}}^{[q+1]} = \left[ \Re \left\{\mathbf{Q}^{\rm H} \mathbf{Q} \right\} + \frac{1}{2} \sigma^2 \mathbf{I}_L \right]^{-1} \Re \left\{\mathbf{Q}^{\rm H} \bar{\bm{y}}  \right\}.
\end{equation}

\subsection{CFO Estimation}

To estimate the CFO, we solve the following unconstrained problem:
\begin{equation}
    \label{phi_objective_function}
    \hat{\phi} \propto \arg \min_{\phi} \mathcal{L}_\phi  \propto \arg \min_{\phi} \Vert \bm{y} - \bm{\mu} \Vert^2.
\end{equation}
To enhance the tractability of (\ref{phi_objective_function}) and {derive} a closed-form solution, {the first-order Taylor approximation is employed}~{\color{black}\cite{liu2024uav, zhang2024joint}}. The $n$-th term on the diagonal of $\bm{\Lambda}_\phi$ can be approximated as
\begin{equation}
    \label{CFO_approximation}
    \left[ \bm{\Lambda}_{\phi} \right]_{n,n} = e^{j 2 \pi \frac{n}{N} \phi} \approx e^{j 2 \pi \frac{n}{N} \hat{\phi}} + \left( \phi - \hat{\phi} \right) \frac{j 2 \pi {\color{black} n}}{N} e^{j 2 \pi \frac{n}{N} \hat{\phi}}.
\end{equation}
Hence, $\hat{\bm{\Lambda}}_{\phi}^{[q+1]}$ can be approximated as
\begin{equation}
\label{Lambda_phi_approximate}
    \hat{\bm{\Lambda}}_{\phi}^{[q+1]}	\approx \hat{\bm{\Lambda}}_\phi^{[q]} +  \left( \phi - \hat{\phi}^{[q]} \right)  \tilde{\bm{\Lambda}}_\phi^{[q]},
\end{equation}
where $\left[ \tilde{\bm{\Lambda}}_{\phi}^{[q]} \right]_{n,n} = \frac{j 2 \pi n}{N} e^{\frac{j 2 \pi n \hat{\phi}^{[q]}}{N}}$.
{\color{black}By defining $x = \phi - \hat{\phi}$, the approximation in (\ref{CFO_approximation}) can be rewritten as
    \begin{equation}
        f(x) = e^{j 2 \pi \frac{n}{N} x} \approx 1 + \frac{j 2 \pi {\color{black} n}}{N} x,
    \end{equation}
    which is essentially the first-order Taylor series expansion of $f(x)$ at $x = 0$, with $x$ being the estimation accuracy of $\phi$. The Lagrange remainder of this Taylor expansion, which reflects the approximation accuracy, can be calculated as
    \begin{align}
        {\rm R}_1(x) &= \frac{f^{''}(\xi)}{2} x^2 
        = - 2 \pi^2 \frac{n^2}{N^2} e^{j 2 \pi \frac{n}{N} \xi} x^2,
    \end{align}
    where $\xi \in (0, x)$. Since $|e^{j 2 \pi \frac{n}{N} \xi}| = 1$, $|{\rm R}_1(x)| \leq 2 \pi^2 \frac{n^2}{N^2} x^2$. The remainder of the first-order Taylor expansion of $f(x)$ is upper bounded, and the upper bound depends on the estimation accuracy of $\phi$. As the estimation accuracy of $\phi$ improves, the Taylor approximation in (\ref{CFO_approximation}) becomes increasingly accurate.}

With (\ref{Lambda_phi_approximate}), $\mathcal{L}_\phi$ in (\ref{phi_objective_function}) can be rewritten as
\begin{equation}
\label{L_phi}
\mathcal{L}_\phi^{[q+1]}  \approx \frac{1}{\sigma^2} \Vert {\bm{y} - \left[ \hat{\bm{\Lambda}}_\phi^{[q]} +  \left( \phi - \hat{\phi}^{[q]} \right)  \tilde{\bm{\Lambda}}_\phi^{[q]}  \right] \bm{d}} \Vert^2,
\end{equation}
where $\bm{d} \triangleq \sqrt{P}  \hat{\bm{\Lambda}}_{\bm{\theta}}^{[q+1]} \mathbf{S} ^ \top \hat{\bm{h}}^{[q]}$.

Upon setting the partial derivative of (\ref{L_phi}) with {regard} to $\phi$ to zero and solving it for $\phi$, we {acquire} the closed-form {solution} of $\hat{\phi}$ at the $(q+1)$-th iteration as
\begin{align}
\label{phi_es}
\hat{\phi}^{[q+1]} =  \hat{\phi}^{[q]} +  \frac{\Re \left\{
    \left( \bm{y} -  \hat{\bm{\Lambda}}_\phi^{[q]}  \bm{d}\right)^{\rm H}  \tilde{\bm{\Lambda}}_\phi^{[q]} \bm{d}
    \right\}}{\bm{d}^{\rm H} \tilde{\bm{\Lambda}}_{\phi}^{[q]{\rm H}} \tilde{\bm{\Lambda}}_\phi^{[q]} \bm{d}}.
\end{align}

\subsection{Position Estimation}
{\color{black}The GD algorithm is a general method for searching maxima or minima in unconstrained optimization problems and has been widely used to solve localization problems in existing studies \cite{liu2022fast, vu2023local, ma2023distributed}. Under the given CFO and PN estimates, the MAP-based problem in (\ref{objective}) is an unconstrained optimization problem. We employ the GD algorithm to estimate UE position.}

Due to the same structures in the respective derivatives of the objective function with {regard} to the three position parameters ${\color{black}d_\text{R}}$, ${\color{black}\phi_\text{az}}$, and ${\color{black}\phi_\text{el}}$, we use the symbol $\xi$ to represent any one of them. The partial derivative of the objective function with {regard} to $\xi$ is given by
\begin{equation}
\frac{\partial \mathcal{L}_\xi }{\partial \xi} = \frac{2}{\sigma^2} \Re \left\{
    \left( P \bm{h}^{\rm H}  \mathbf{S}^* \mathbf{S}^\top - \sqrt{P} \bm{y}^{\rm H} \bm{\Lambda}_{\bm{\theta}} \bm{\Lambda}_\phi \mathbf{S}^\top \right) \frac{\partial \bm{h}}{\partial \xi}
\right\},
\end{equation}
where $\frac{\partial \bm{h}}{\partial \xi}$ {is} calculated {by}
\begin{align}
    \frac{\partial \bm{h}}{\partial \xi}[n] &= \frac{\partial \bm{h}[n]}{\partial \xi} = - \sum_{r=0}^{{\color{black} N_{\text{R}}-1}} {\color{black}\sqrt{\rho_{{\text{aR}}}^{n,r}}} {\color{black}\sqrt{\rho_{{\text{Ru}}}^{n,r}}} \bm{w}[r] \notag \\
    &\times e^{-j 2 \pi \frac{n B }{N}({\tau_{{\color{black} \text{a}}r}} + {\color{black} \tau_{r\text{u}}})} \left( \frac{1}{{\color{black} d_{r\text{u}}}} + j 2 \pi \frac{n B }{N c} \right) \frac{\partial {\color{black} d_{r\text{u}}}}{\partial \xi},
\end{align}
with $\frac{\partial {\color{black} d_{r\text{u}}}}{\partial \xi}$, $\xi={\color{black}d_\text{R}}, {\color{black}\phi_\text{az}}, {\color{black}\phi_\text{el}}$, being
\begin{align}
    \frac{\partial {\color{black} d_{r\text{u}}}}{\partial {\color{black}d_\text{R}}} &= \frac{1}{{\color{black}{\color{black} d_{r\text{u}}}}}  \left[ ({\color{black}d_\text{R}} \sin {\color{black}\phi_\text{el}} \cos {\color{black}\phi_\text{az}} - x_r) \sin {\color{black}\phi_\text{el}} \cos {\color{black}\phi_\text{az}} \notag \right. \\
    &\left. \quad+ ({\color{black}d_\text{R}} \sin {\color{black}\phi_\text{el}} \sin {\color{black}\phi_\text{az}} - y_r) \sin {\color{black}\phi_\text{el}} \sin {\color{black}\phi_\text{az}} \notag \right. \\
    &\left. \quad+ ({\color{black}d_\text{R}} \cos {\color{black}\phi_\text{el}} - z_r) \cos {\color{black}\phi_\text{el}} \right],
\end{align}
\begin{align}
    \frac{\partial {\color{black} d_{r\text{u}}}}{\partial {\color{black}\phi_\text{az}}} &= \frac{1}{{\color{black}{\color{black} d_{r\text{u}}}}}  \left[
	({\color{black}d_\text{R}} \sin {\color{black}\phi_\text{el}} \cos {\color{black}\phi_\text{az}} - x_r) (- \sin {\color{black}\phi_\text{az}} {\color{black}d_\text{R}} \sin {\color{black}\phi_\text{el}}) \notag \right. \\
    & \left. \quad+ ({\color{black}d_\text{R}} \sin {\color{black}\phi_\text{el}} \sin {\color{black}\phi_\text{az}} - y_r) {\color{black}d_\text{R}} \sin {\color{black}\phi_\text{el}} \cos {\color{black}\phi_\text{az}} \right],
\end{align}
and
\begin{align}
    \frac{\partial {\color{black} d_{r\text{u}}}}{\partial {\color{black}\phi_\text{el}}} &= \frac{1}{{\color{black}{\color{black} d_{r\text{u}}}}}  \left[
	({\color{black}d_\text{R}} \sin {\color{black}\phi_\text{el}} \cos {\color{black}\phi_\text{az}} - x_r) {\color{black}d_\text{R}} \cos {\color{black}\phi_\text{az}} \cos {\color{black}\phi_\text{el}}  \notag \right. \\
    & \left. \quad+ ({\color{black}d_\text{R}} \sin {\color{black}\phi_\text{el}} \sin {\color{black}\phi_\text{az}} - y_r) {\color{black}d_\text{R}} \sin {\color{black}\phi_\text{az}} \cos {\color{black}\phi_\text{el}} \notag \right. \\
    & \left. \quad+ ({\color{black}d_\text{R}} \cos {\color{black}\phi_\text{el}} - z_r)  (-{\color{black}d_\text{R}}  \sin {\color{black}\phi_\text{el}})\right].
\end{align}

\subsection{Overall Algorithm}

The proposed joint estimation of the CFO, PN and UE position is summarized in Algorithm \ref{alg:A}. Specifically, we initialize the parameters, including the initial estimation of the normalized CFO $\hat{\phi}$ and UE position $\hat{\bm{p}}_{\color{black}\text{u}}$, the step size $k_{\color{black} \text{p}}$, the convergence accuracy requirements for outer and inner loops, $\epsilon_{\color{black} \text{O}}$ and $\epsilon_{\color{black} \text{I}}$, and the maximum numbers of iterations for the outer and inner loops, $N_{\color{black} \text{O} }$ and $N_{\color{black} \text{I} }$. {\color{black}Suppose that the range of the normalized CFO $\phi$ is known at the UE, and the initial $\hat{\phi}$ is chosen randomly within this range. The initial estimation of the UE position, $\hat{\bm{p}}_{\color{black}\text{u}}$, is obtained using the state of the art proposed in \cite{luan2021phase}. The other initial parameters are hyperparameters, and can be fine-tuned to balance both convergence speed and convergence accuracy.}

Then, after transforming the coordinates into the polar coordinates, we estimate the PN, CFO, and UE position alternately. Closed-form expressions are utilized to update the PN and CFO estimates in each iteration, while the GD algorithm is employed to estimate the UE position. This repeats until {either the objective function $\mathcal{L}$ converges} or the maximum number of iterations is reached. Notably, this AO-based algorithm takes the RIS phase shifts, i.e., $\bm{w}$, as input, which can affect the algorithm performance significantly. The {configuration} of the RIS phase shifts, $\bm{w}$, will be discussed in Section \ref{section: RIS_design}.

In Algorithm \ref{alg:A}, the objective function is expected to be non-increasing throughout iterations when updating the CFO and PN parameters because the approximate functions in (\ref{L_theta}) and~(\ref{L_phi}) show convexity with respect to $\bm{\eta}$ and $\phi$, respectively. In Steps~9, 10 and 11, with the UE position estimated in the previous iteration as input, the GD algorithm can ensure a lower objective function. Since the posterior probability is no larger than $1$, the negative logarithmic function $\mathcal{L}(\bm{\lambda})$ is non-negative and {therefore} lower bounded. Algorithm \ref{alg:A} converges. 

The computational complexity for updating $\bm{\eta}$, $\bm{\theta}$, $\phi$, and position parameters ${\color{black}d_\text{R}}$, ${\color{black}\phi_\text{az}}$, and ${\color{black}\phi_\text{el}}$ in Algorithm \ref{alg:A} are $\mathcal{O}(L^3+N^2+NL^2)$, $\mathcal{O}(NL)$, $\mathcal{O}(N^2)$, and {$\mathcal{O}(\max \left\{N^3, {\color{black} N_{\text{R}}}\right\})$}, respectively. Suppose the inner GD algorithm for updating the UE position converges after $K_{\color{black} \text{I} }$ iterations. {Consequently}, the computational complexity of Algorithm \ref{alg:A} is {$\mathcal{O}(\max \left\{L^3+N^2+NL^2, K_{\color{black} \text{I} }N^3, K_{\color{black} \text{I} }{\color{black} N_{\text{R}}} \right\})$} per iteration. {\color{black}By contrast, the complexity of the UE localization method in \cite{luan2021phase} is $\mathcal{O}(\max (K_\text{I} N^3, K_\text{I} N_\text{R}))$ per iteration. Nevertheless, by adopting an AO-based approach to jointly estimate the UE position, CFO, and PN, our algorithm can achieve up to a two-order-of-magnitude improvement in localization accuracy under high SNR conditions, compared to the method developed in \cite{luan2021phase}, as demonstrated numerically in Section \ref{section:results}-C}.

\begin{algorithm}[t]
\caption{Proposed iterative joint estimation algorithm.}  
\label{alg:A}  
\begin{algorithmic}[1]

\STATE {\textbf{Input: } $\bm{w}$;}
\STATE \textbf{Initialization:} $\hat{\phi}$, $\hat{\bm{p}}_{\color{black}\text{u}}$, $k_{\color{black} \text{p} }$, $\epsilon_{\color{black} \text{O}}$, $\epsilon_{\color{black} \text{I}}$, $N_{\color{black} \text{O} }$, $N_{\color{black} \text{I} }$;
\STATE {Calculate polar coordinate of UE using (\ref{eq:cartesian_2_polar});}
\REPEAT   
   \STATE Given fixed $\hat{\bm{p}}_{\color{black}\text{u}}$, update $\hat{\bm{\eta}}$ with $\hat{\phi}$ using (\ref{eta_es});
    \STATE Update $\hat{\bm{\theta}}$ using (\ref{eta_2_theta});
    \STATE Given fixed $\hat{\bm{p}}_{\color{black}\text{u}}$, update $\hat{\phi}$ with $\hat{\bm{\theta}}$ using (\ref{phi_es});
    \REPEAT
        \STATE $\hat{d}_{\color{black}\text{R}} \leftarrow \hat{d}_{\color{black}\text{R}} - \frac{\partial \mathcal{L}_\xi}{\partial {\color{black}d_\text{R}}} k_{\color{black} \text{p} }$;
        \STATE $\hat{\phi}_{\color{black}\text{az}} \leftarrow \hat{\phi}_{\color{black}\text{az}} - \frac{\partial \mathcal{L}_\xi}{\partial {\color{black}\phi_\text{az}}} k_{\color{black} \text{p} }$;
        \STATE $\hat{\phi}_{\color{black}\text{el}} \leftarrow \hat{\phi}_{\color{black}\text{el}} - \frac{\partial \mathcal{L}_\xi}{\partial {\color{black}\phi_\text{el}}} k_{\color{black} \text{p} }$;
    \UNTIL{$\mathcal{L}^{[p]} - \mathcal{L}^{[p+1]} \leq \epsilon_{\color{black}\text{I}}$ after the $p$-th iteration} or $p \textgreater N_{\color{black} \text{I} }$.
\UNTIL{$\mathcal{L}^{[q]} - \mathcal{L}^{[q+1]} \leq \epsilon_{\color{black}\text{O}}$ after the $q$-th iteration or $q \textgreater N_{\color{black} \text{O} }$.}

\STATE Calculate Cartesian coordinate of UE using (\ref{eq:polar_2_cartesian});
\STATE \textbf{Output:} $\hat{\phi}$, $\hat{\bm{\theta}}$, $\hat{\bm{p}}_{\color{black}\text{u}}$.

\end{algorithmic}
\end{algorithm}  

\subsection{Performance Metrics}
\label{PerformanceMetrics}
We specify the performance metrics for all estimation parameters in $\bm{\lambda}$ {for the sake of evaluating Algorithm \ref{alg:A}}. For the position estimation, we take \emph{root mean square error} (RMSE) as the performance {index}. For any testing UE position $\bm{p}_{\color{black}\text{u}}$, ${\rm RMSE}_{\color{black}\text{u}}$ can be expressed as
\begin{equation}
\label{RMSE}
{\rm RMSE}_{\color{black}\text{u}} = \sqrt{\frac{1}{M_{\color{black} \text{e}}} {\color{black}\sum_{m=0}^{M_{\text{e}} - 1}} \Vert \hat{\bm{p}}_{{\color{black}\text{u}},m} - \bm{p}_{\color{black}\text{u}} \Vert ^2 },
\end{equation}
where $M_{\color{black} \text{e}}$ is the number of Monte Carlo tests, and $\hat{\bm{p}}_{{\color{black}\text{u}},m}$ {denotes} the estimation of the UE position in the $m$-th trial.

For the CFO and PN estimation, it can be straightforwardly observed from (\ref{receivedSignalWithoutHNLoS}) that the received signal $\bm{y}$ does not change under a common phase rotation $\epsilon$, i.e.,
\begin{equation}
    \label{ambiguity}
    \hat{\phi} \rightarrow \phi - \epsilon \quad \text{and } \quad \hat{\bm{\theta}} \rightarrow \bm{\theta} + \bm{\epsilon},
\end{equation}
where $\bm{\epsilon}[n] = \frac{2 \pi n \epsilon}{N}$. Here, (\ref{ambiguity}) indicates that there exists ambiguity in estimating the CFO and PN. The ambiguity leads to difficulties in achieving high estimation accuracy for individual parameters of the CFO and PN. Nevertheless, a high-precision joint estimation of CFO and PN is achievable, which is beneficial for estimating the UE position and will be illustrated in section \ref{section:results}.

To evaluate the joint estimation performance, we utilize a joint MSE measure, which can be calculated as
\begin{equation}
\label{jointMSE}
    {\rm MSE}_{\phi, \bm{\theta}} = \mathbb{E}_{\phi, \bm{\theta}}  \left( \Vert \bm{\underline{\gamma}} - \hat{\bm{\underline{\gamma}}}  \Vert^2 \right),
\end{equation}
where $\bm{\underline{\gamma}} = \bm{\gamma} - \bm{\gamma}[0] \bm{1}$, $\hat{\bm{\underline{\gamma}}} = \hat{\bm{\gamma}} - \hat{\bm{\gamma}}[0] \bm{1}$, $\bm{\gamma} = [\bm{\gamma}[0], \bm{\gamma}[1], \cdots, \bm{\gamma}[N-1]]^\top$, $\bm{\gamma}[n] = \bm{\theta}[n] + \frac{2 \pi n \phi}{N}$, $\hat{\bm{\gamma}}= [\hat{\bm{\gamma}}[0], \hat{\bm{\gamma}}[1], \cdots, \hat{\bm{\gamma}}[N-1]]^\top$, and $\hat{\bm{\gamma}}[n] = \hat{\bm{\theta}}[n] + \frac{2 \pi n \hat{\phi}}{N}$. {\color{black}By focusing on joint CFO-PN estimation rather than individual variables, we ensure a more comprehensive performance evaluation. Subtracting the initial elements of $\bm{\gamma}$ and $\hat{\bm{\gamma}}$, specifically the actual PN $\bm{\theta}[0]$ and its estimate $\hat{\bm{\theta}}[0]$ at time $0$, helps mitigate the influence of the initial PN $\bm{\theta}[0]$ (and its estimate $\hat{\bm{\theta}}[0]$) on the MSE analysis and resulting in a more precise evaluation of the joint CFO-PN estimation.}

\section{Hybrid Cram\'{e}r-Rao Lower Bound Analysis}
\label{section: HCRLB}
{HCRLB for the joint estimation of the CFO, PN, and UE position is derived in this section.} This helps quantify the performance lower bound of the proposed AO-based estimation algorithm, and serves as the objective function for optimizing the RIS phase shifts, $\bm{w}$, in the next section. {\color{black}HCRLB is an extension of the traditional CRLB, suitable for estimation problems with both deterministic and random unknown variables. Unlike CRLB, HCRLB exploits the prior distribution of the unknown random variables to improve analysis. Since we estimate the deterministic variables, including CFO and UE position, as well as the random variable, i.e., PN, analyzing the HCRLB allows for more accurate estimation error lower bounds for these variables.}

{\color{black}The HCRLB for the estimate of $\bm{\lambda}$ is~\cite[Eq.~(4.610)]{van2004detection}}
\begin{equation}
    \mathbb{E}_{\bm{y}, \bm{\theta} \mid \phi, {\color{black}d_\text{R}}, {\color{black}\phi_\text{az}}, {\color{black}\phi_\text{el}} } \left[ 
        (\bm{\lambda} - \hat{\bm{\lambda}}) (\bm{\lambda} - \hat{\bm{\lambda}}) ^ \top
    \right] \succeq \mathbf{B}^{\rm -1},
\end{equation}
where $\mathbf{B}$ is the \emph{Bayesian information matrix} (BIM), which can be given by~\cite[Eq.~(4.611)]{van2004detection}
\begin{equation}
\label{BIM}
    \mathbf{B} = \mathbb{E}_{\bm{\theta}} \left[ \bm{\mathrm{FIM}}(\bm{y}; \bm{\lambda}) \right] +  \mathbb{E}_{\bm{\theta}} \left[ - \frac{\partial  }{\partial \bm{\lambda}} \left(\frac{\partial \log p(\bm{\theta}) }{\partial \bm{\lambda}}\right)^\top   \right],
\end{equation}
where $\bm{\mathrm{FIM}}(\bm{y}; \bm{\lambda})$ stands for the \emph{Fisher's information matrix} (FIM) for $\bm{\lambda}$ under the observation of received signal $\bm{y}$ at the UE, and $\mathbb{E}_{\bm{\theta}} \left[ - \frac{\partial  }{\partial \bm{\lambda}} \left(\frac{\partial \log p(\bm{\theta}) }{\partial \bm{\lambda}}\right)^\top   \right]$ is the prior information matrix. Interested readers may refer to \cite{van2004detection} for more details.

\subsection{Derivation of $\mathbb{E}_{\bm{\theta}} \left[ \bm{\mathrm{FIM}}(\bm{y}; \bm{\lambda}) \right]$}

To derive $\mathbb{E}_{\bm{\theta}} \left[ \bm{\mathrm{FIM}}(\bm{y}; \bm{\lambda}) \right]$ in (\ref{BIM}), we first analyze the FIM for $\bm{\lambda}$, as given by
\begin{equation}
\label{FIM}
    \left[\bm{\mathrm{FIM}}(\bm{y}; \bm{\lambda})  \right]_{i, j} = \frac{2}{\sigma^2} \Re \left\{\frac{\partial \bm{\mu}{\rm ^{\rm H}}}{\partial \lambda_i} \frac{\partial \bm{\mu}}{\partial \lambda_j} \right\}.
\end{equation}

To calculate (\ref{FIM}), the first-order derivative of $\bm{\mu}$ with respect to each element in vector $\bm{\lambda}$ is derived as follows.
\begin{subequations} \label{eq:mu_partials}
\begin{align}
\frac{\partial \bm{\mu}}{\partial \phi} &= \sqrt{P} \bm{\Lambda} \bm{\Lambda}_\phi \bm{\Lambda}_{\bm{\theta}} \mathbf{S} ^ \top  \bm{h};  \\
\frac{\partial \bm{\mu}}{\partial \bm{\theta}[n]} &= \sqrt{P} {\rm Diag}(\bm{a}[n]) \bm{\Lambda}_\phi \mathbf{S} ^ \top  \bm{h}; \\
\frac{\partial \bm{\mu}}{\partial {\color{black}d_\text{R}}} &= \sqrt{P} \bm{\Lambda}_\phi \bm{\Lambda}_{\bm{\theta}} \mathbf{S} ^ \top  \frac{\partial \bm{h} }{\partial {\color{black}d_\text{R}}}; \\
\frac{\partial \bm{\mu}}{\partial {\color{black}\phi_\text{az}}} &=  \sqrt{P} \bm{\Lambda}_\phi \bm{\Lambda}_{\bm{\theta}} \mathbf{S} ^ \top  \frac{\partial \bm{h}}{\partial {\color{black}\phi_\text{az}}};  \\
\frac{\partial \bm{\mu}}{\partial {\color{black}\phi_\text{el}}} &=  \sqrt{P} \bm{\Lambda}_\phi \bm{\Lambda}_{\bm{\theta}} \mathbf{S} ^ \top  \frac{\partial \bm{h}}{\partial {\color{black}\phi_\text{el}}}, 
\end{align}
\end{subequations}
where $\bm{\Lambda}$ is a diagnal matrix, $\left[\bm{\Lambda}\right]_{n, n} = \frac{j 2 \pi n}{N}$, and $\bm{a}[n] = [\bm{0}_{1 \times n}, j e^{j \bm{\theta}[n]}, \bm{0}_{1 \times (N-n-1)}]^\top$.

It is not difficult to see that in (\ref{eq:mu_partials}), only $\bm{\Lambda}_\phi$, $\bm{\Lambda}_{\bm{\theta}}$ and ${\rm Diag}(\bm{a}[n])$ are related to $\phi$ and $\bm{\theta}$, which are all diagonal matrices. Additionally, both $\phi$ and $\bm{\theta}$ appear in the exponent of diagonal elements in these matrices. Hence, they can be suppressed when calculating $\frac{\partial \bm{\mu}{\rm ^{\rm H}}}{\partial \lambda_i} \frac{\partial \bm{\mu}}{\partial \lambda_j}$ in (\ref{FIM}). In other words, the calculated FIM does not rely on the parameter $\bm{\theta}$, i.e.,
\begin{equation}
    \mathbb{E}_{\bm{\theta}} \left[ \bm{\mathrm{FIM}}(\bm{y}; \bm{\lambda}) \right] = \bm{\mathrm{FIM}}(\bm{y}; \bm{\lambda}).
\end{equation}

Take $\frac{\partial \bm{\mu}^{\rm H}}{\partial {\color{black}d_\text{R}}} \frac{\partial \bm{\mu}}{\partial {\color{black}d_\text{R}}}$ as an example. We have 
\begin{align}
    \frac{\partial \bm{\mu}^{\rm H}}{\partial {\color{black}d_\text{R}}} \frac{\partial \bm{\mu}}{\partial {\color{black}d_\text{R}}} &= \sqrt{P}  \frac{\partial \bm{h}^{\rm H} }{\partial {\color{black}d_\text{R}}} \mathbf{S}^* \bm{\Lambda}_{\bm{\theta}}^{\rm{H}} \bm{\Lambda}_\phi^{\rm H} \sqrt{P} \bm{\Lambda}_\phi \bm{\Lambda}_{\bm{\theta}} \mathbf{S} ^ \top  \frac{\partial \bm{h} }{\partial {\color{black}d_\text{R}}} \\ \notag
    &= P \frac{\partial \bm{h}^{\rm H} }{\partial {\color{black}d_\text{R}}} \mathbf{S}^* \mathbf{S} ^ \top  \frac{\partial \bm{h} }{\partial {\color{black}d_\text{R}}},
\end{align}
which is unrelated to $\phi$ or $\bm{\theta}$, since $\bm{\Lambda}_\phi^{\rm H} \bm{\Lambda}_\phi = \bm{\Lambda}_{\bm{\theta}}^{\rm{H}} \bm{\Lambda}_{\bm{\theta}} = \mathbf{I}_N$. This can be extended to $\frac{\partial \bm{\mu}{\rm ^{\rm H}}}{\partial \lambda_i} \frac{\partial \bm{\mu}}{\partial \lambda_j}, \forall i, j$, and is omitted here. 

\subsection{Derivation of $\mathbb{E}_{\bm{\theta}} \left[ - \frac{\partial  }{\partial \bm{\lambda}} \left(\frac{\partial \log p(\bm{\theta}) }{\partial \bm{\lambda}}\right)^\top \right]$}

Since $\bm{\theta} \sim \mathcal{N}(\bm{0}_{N \times 1}, \bm{\Psi})$, we can calculate $\log p(\bm{\theta})$ as $\log p(\bm{\theta}) = - \frac{1}{2} \bm{\theta}^\top \bm{\Psi}^{-1} \bm{\theta} - \frac{N}{2}\log(2 \pi) - \frac{1}{2}\log \det \bm{\Psi }$, and its partial derivative with respect to $\bm{\lambda}$ can be calculated as
\begin{equation}
	\frac{\partial \log p(\bm{\theta})}{\partial \bm{\lambda}} = \left[ 0, - (\bm{\Psi}^{-1} \bm{\theta})^\top, 0, 0, 0 \right]^\top.
\end{equation}
Subsequently, $\mathbb{E}_{\bm{\theta}} \left[ - \frac{\partial  }{\partial \bm{\lambda}} \left(\frac{\partial \log p(\bm{\theta}) }{\partial \bm{\lambda}}\right)^\top \right]$ can be expressed as
\begin{align}
    \mathbb{E}_{\bm{\theta}} \!\left[ - \!\frac{\partial  }{\partial \bm{\lambda}} \left(\frac{\partial \log p(\bm{\theta}) }{\partial \bm{\lambda}}\right)^\top \right] &=\!  - \frac{\partial }{\partial \bm{\lambda}} \left(\!\frac{\partial \log p(\bm{\theta}) }{\partial \bm{\lambda}}\!\right)^\top \notag \\
    = & \mathrm{Blkdiag}(0, \bm{\Psi}^{-1} , \bm{0}_{3 \times 3}).
\end{align}

\subsection{Derivation of Transformed HCRLB}

So far, we have derived the expression for $\mathbf{B}$, {\color{black}whose inverse} is an HCRLB matrix that gives analytical estimation accuracy lower bounds for the parameters in $\bm{\lambda}$. However, when assessing the proposed estimation algorithm, we adopt the RMSE and CFO-PN joint MSE metrics introduced in Section~\ref{PerformanceMetrics}. Hence, we define a new parameter vector $\bm{\alpha} = \bm{g}(\bm{\lambda}) \triangleq \left[ \bm{\underline{\gamma}},  x_{\color{black}\text{u}}, y_{\color{black}\text{u}}, z_{\color{black}\text{u}} \right]^\top$, corresponding to the two metrics. Then, $\mathbf{B}$ can be transformed into the $\bm{\alpha}$-related HCRLB matrix~\cite{kay1993fundamentals}
\begin{equation}
\label{HCRLB}
    \mathbf{HCRLB}(\bm{p}_{\color{black}\text{u}}, \bm{w}) = \bm{\Xi}(\bm{p}_{\color{black}\text{u}}) \mathbf{B}^{-1}(\bm{p}_{\color{black}\text{u}}, \bm{w}) \bm{\Xi}^\top({\bm{p}_{\color{black}\text{u}}}),
\end{equation}
where, for any UE position $\bm{p}_{\color{black}\text{u}}$, $\mathbf{\Xi}(\bm{p}_{\color{black}\text{u}})$ is the transition matrix defined as 
\begin{equation}
    \bm{\Xi}(\bm{p}_{\color{black}\text{u}}) = \frac{\partial \bm{g}(\bm{\lambda})}{\partial \bm{\lambda}} = {\rm Blkdiag} \left(\bm{\Xi}_{1}, \bm{\Xi}_{2}(\bm{p}_{\color{black}\text{u}})\right),
\end{equation}
with $\bm{\Xi}_{1}$ and $\bm{\Xi}_{2}(\bm{p}_{\color{black}\text{u}})$ given by (\ref{xi_1}) and (\ref{xi_2}), respectively.
\begin{figure*}[!t]
\normalsize
\begin{align} \label{xi_1}
    \bm{\Xi}_{1} = \begin{bmatrix}
        \frac{\partial \bm{\underline{\gamma}}[0]}{\partial \phi}  &  \frac{\partial \bm{\underline{\gamma}}[0]}{\partial \bm{\theta}[0]} & \cdots & \frac{\partial \bm{\underline{\gamma}}[0]}{\partial \bm{\theta}[N-1]}  \\
        \vdots  & \vdots & \ddots & \vdots \\
        \frac{\partial \bm{\underline{\gamma}}[N-1]}{\partial \phi} & \frac{\partial \bm{\underline{\gamma}}[N-1]}{\partial \bm{\theta}[0]} & \cdots & \frac{\partial \bm{\underline{\gamma}}[N-1]}{\partial \bm{\theta}[N-1]} 
    \end{bmatrix} = \begin{bmatrix}
		0 & 0 & 0 & \cdots & 0 \\
		\frac{2 \pi}{N} & -1 & 1 & \cdots & 0 \\
		\vdots & \vdots & \vdots & \ddots & \vdots \\
		\frac{2 \pi (N-1)}{N} & -1 & 0 & \cdots & 1\\		
	\end{bmatrix} \in \mathbb{R}^{N \times (N+1)},
\end{align}
\begin{align} \label{xi_2}
    \bm{\Xi}_{2}(\bm{p}_{\color{black}\text{u}}) =
	\begin{bmatrix}
		\frac{\partial x_{\color{black}\text{u}}}{\partial {\color{black}d_\text{R}}} &   \frac{\partial x_{\color{black}\text{u}}}{\partial {\color{black}\phi_\text{az}}} & \frac{\partial x_{\color{black}\text{u}}}{\partial {\color{black}\phi_\text{el}}}\\
		\frac{\partial y_{\color{black}\text{u}}}{\partial {\color{black}d_\text{R}}} &  \frac{\partial y_{\color{black}\text{u}}}{\partial {\color{black}\phi_\text{az}}} & \frac{\partial y_{\color{black}\text{u}}}{\partial {\color{black}\phi_\text{el}}}\\
		 \frac{\partial z_{\color{black}\text{u}}}{\partial {\color{black}d_\text{R}}} &   \frac{\partial z_{\color{black}\text{u}}}{\partial {\color{black}\phi_\text{az}}} &  \frac{\partial z_{\color{black}\text{u}}}{\partial {\color{black}\phi_\text{el}}}\\
	\end{bmatrix} = \begin{bmatrix}
		\sin {\color{black}\phi_\text{el}} \cos {\color{black}\phi_\text{az}} & -{\color{black}d_\text{R}} \sin {\color{black}\phi_\text{el}} \sin {\color{black}\phi_\text{az}}  &  {\color{black}d_\text{R}} \cos {\color{black}\phi_\text{az}} \cos {\color{black}\phi_\text{el}} \\
		\sin {\color{black}\phi_\text{el}} \sin {\color{black}\phi_\text{az}} & {\color{black}d_\text{R}} \sin {\color{black}\phi_\text{el}} \cos {\color{black}\phi_\text{az}} & {\color{black}d_\text{R}} \sin {\color{black}\phi_\text{az}} \cos {\color{black}\phi_\text{el}} \\
		\cos {\color{black}\phi_\text{el}} & 0 & -{\color{black}d_\text{R}} \sin {\color{black}\phi_\text{el}}
	\end{bmatrix} \in \mathbb{R}^{3 \times 3}.
\end{align}
\hrulefill
\vspace*{-5pt}
\end{figure*}

After obtaining $\mathbf{HCRLB}(\bm{p}_{\color{black}\text{u}}, \bm{w})$, the \emph{position error bound} (${\rm PEB}$) for the UE position estimation and $\rm HCRLB_{CFO-PN}$ for the CFO-PN joint estimation can be given by
\begin{equation}
\label{PEB}
	 {\rm PEB} = \sqrt{{\rm tr}([\mathbf{HCRLB}(\bm{p}_{\color{black}\text{u}}, \bm{w})]_{N: N+2, N: N+2})};
\end{equation}
\begin{equation}
     {\rm HCRLB_{CFO-PN}} = {\rm tr}([\mathbf{HCRLB}(\bm{p}_{\color{black}\text{u}}, \bm{w})]_{0: N-1, 0: N-1}).
\end{equation}
Both ${\rm PEB}$ and ${\rm HCRLB_{CFO-PN}}$ depend on the UE's {position}, $\bm{p}_{\color{black}\text{u}}$, and the RIS phase shifts, $\bm{w}$.

\section{RIS Configuration}
\label{section: RIS_design}
In this section, we {configure} the phase shifts $\bm{w}$ of the RIS to minimize the analytical lower bound of {\color{black}the localization accuracy, the PEB derived in Section \ref{section: HCRLB}. The configuration effectively enhances the performance of the algorithm proposed in Section~III, and is applied before the localization process.}

According to (\ref{PEB}), the 
PEB depends on the UE position, which indicates that minimizing the PEB at a specific UE position requires knowledge of that position. However, this is not possible in practice. To address this issue and enhance the {algorithm's} robustness, {\color{black}we specify the \emph{area of interest} (AOI), according to the prior knowledge and practical requirements. The AOI represents the region in which we anticipate the UE of interest is situated and the high-precision localization of the UE is expected. The design objective of the RIS configuration is to achieve high expected accuracy for the localization of UEs located within the AOI, i.e., by minimizing the average PEB within the AOI.} Specifically, we {\color{black}uniformly and randomly} sample {\color{black}$S$} positions within the AOI{\color{black}, denoted as $\bm{p}_s, s = 0, 1, \cdots, S-1$,} and {\color{black}evaluate} their average PEB.

{\color{black}This problem is transformed into a convex SDP form using the SDR method and the Schur complement. Unlike traditional approaches that rely on the classical FIM, the objective function represented by the average PEB is derived from the inverse of the BIM, $\mathbf{B}$. To establish the convexity of the transformed problem, we analytically reveal the linear relationship between $\mathbf{B}$ and the optimization variable  $\mathbf{W}$.}

Accordingly, the problem of interest is formulated as
\begin{align}
\label{originOptimizationProblem}
    &\min_{\bm{w}} \quad \frac{1}{\color{black}S} {\color{black}\sum_{s=0}^{S-1}} {\rm PEB}(\bm{p}_{\color{black}s}, \bm{w}) \notag \\
    & {\color{black}\rm s.t.}  \quad  | \bm{w}[r] | = 1,  r={\color{black}0, 1, \cdots, }{\color{black} N_{\text{R}} - 1}.
\end{align}
Problem (\ref{originOptimizationProblem}) is {not convex} because of the non-convexity of the objective function with respect to the vector $\bm{w}$, as well as the unit-module constraint.

To convexify (\ref{originOptimizationProblem}), we first define $\mathbf{W} \triangleq \bm{\bar{w}} \bm{\bar{w}}^{\rm H}$, where $\bm{\bar{w}} \triangleq \bm{w}^*$, which satisfies $\mathbf{W} \succcurlyeq 0$ and ${\rm rank}(\mathbf{W}) = 1$. Since the rank-1 constraint is non-convex, we employ the SDR method to relax this constraint. Consequently, the problem in (\ref{originOptimizationProblem}) can be transformed into
\begin{align}
    \label{problem2}
    & \min_{\mathbf{W}} \quad \frac{1}{\color{black}S} {\color{black}\sum_{s=0}^{S-1}} {\rm PEB}(\bm{p}_{\color{black}s}, \mathbf{W}) \notag \\
    & \begin{array}{ll}
        {\color{black}\rm s.t.}  & \left[ \mathbf{W} \right]_{r, r} = 1, r={\color{black}0, 1}, \cdots, {\color{black} N_{\text{R}}-1}, \\
        & \mathbf{W} \succcurlyeq 0.
    \end{array}
\end{align}

According to (\ref{HCRLB}) and (\ref{PEB}), the {optimization} problem in (\ref{problem2}) {remains} non-convex because of the complex objective function. To transform (\ref{problem2}) into the convex SDP form, for each calculated PEB at a particular position {\color{black}$\bm{p}_{s}$}, we define an auxiliary matrix $\mathbf{Z}_{\color{black}s}$ that satisfies
\begin{equation}
    \label{AuxiliaryMatrix}
    \mathbf{Z}_{\color{black}s} \succcurlyeq \tilde{\bm{\Xi}}(\bm{p}_{\color{black}s}) \mathbf{B}^{-1}(\bm{p}_{\color{black}s}, \mathbf{W}) \tilde{\bm{\Xi}}^\top(\bm{p}_{\color{black}s}),
\end{equation}
where $\tilde{\bm{\Xi}}(\bm{p}_{\color{black}s}) \triangleq \mathbf{T} \mathbf{\Xi}(\bm{p}_{\color{black}s})$, and $\mathbf{T} \triangleq [\bm{0}_{N \times N} \quad \mathbf{I}_{3}] \in \mathbb{R}^{(N+3) \times (N+3)}$ can extract the PEB from the HCRLB. According to the property of the Schur complement and the positive semidefiniteness of $\mathbf{B}(\bm{p}_{\color{black}s}, \mathbf{W})$, (\ref{AuxiliaryMatrix}) can be transformed to
\begin{equation}
    \begin{bmatrix}
        \mathbf{Z}_{\color{black}s} & \tilde{\bm{\Xi}}(\bm{p}_{\color{black}s}) \\
        \tilde{\bm{\Xi}}^\top(\bm{p}_{\color{black}s}) & \mathbf{B}(\bm{p}_{\color{black}s}, \mathbf{W}) 
    \end{bmatrix} \succcurlyeq 0.
\end{equation}

{Thus, }the problem in (\ref{originOptimizationProblem}) can be converted to
\begin{align}
\label{finalOptimizationProblem}
    & \min_{\mathbf{W}, \mathbf{Z}_{\color{black}0}, \cdots, \mathbf{Z}_{\color{black}S-1}} \quad \frac{1}{\color{black}S} {\color{black}\sum_{s=0}^{S-1}} {\rm tr}(\mathbf{Z}_{\color{black}s}) \notag \\
    & \begin{array}{ll}
        {\color{black}\rm s.t.}  & \left[ \mathbf{W} \right]_{r, r} = 1, r = 0, {\color{black}1,} \cdots, {\color{black} N_{\text{R}} - 1}, \\
        & \mathbf{W} \succcurlyeq 0, \\
        & \displaystyle  \frac{1}{\color{black}S} {\color{black}\sum_{s=0}^{S-1}}
        \begin{bmatrix}
            \mathbf{Z}_{\color{black}s} & \tilde{\bm{\Xi}}(\bm{p}_{\color{black}s}) \\
            \tilde{\bm{\Xi}}^\top(\bm{p}_{\color{black}s}) & \mathbf{B}(\bm{p}_{\color{black}s}, \mathbf{W}) 
        \end{bmatrix} \succcurlyeq 0,
    \end{array}
\end{align}
where, according to the relationship in (\ref{AuxiliaryMatrix}) and the calculation of PEB in (\ref{PEB}), minimizing the objective function results in minimizing the average square of PEB, which also minimizes the initial objective function, the average PEB.

\smallskip
\smallskip
\noindent \textbf{Proposition 1.} $\mathbf{B}(\bm{p}_{\color{black}s}, \mathbf{W})$ is linear to $\mathbf{W}$.
\smallskip

\noindent \emph{Proof. } According to (\ref{BIM}), only $\bm{\mathrm{FIM}}(\bm{y}; \bm{\lambda})$ depends on $\bm{w}$ in~$\mathbf{B}$. Hence, we only need to prove the linearity of $\bm{\mathrm{FIM}}(\bm{y}; \bm{\lambda})$ to $\mathbf{W}$, which can be rewritten as
\begin{equation}
\label{FIM_summation}
    \left[ 	\bm{\mathrm{FIM}}(\bm{y}; \bm{\lambda})  \right]_{i, j} = \frac{2}{\sigma^2}  \sum_{n=0}^{N-1}  \Re \left\{\frac{\partial \bm{\mu}[n]^*} {\partial \lambda_i} \frac{\partial \bm{\mu}[n]}{\partial \lambda_j} \right\}.
\end{equation}
To derive the explicit expressions of $\frac{\partial \bm{\mu}[n]^*}{\partial \lambda_i}, \forall i$ and $\frac{\partial \bm{\mu}[n]}{\partial \lambda_j}, \forall j$ with respect to $\bar{\bm{w}}$, we rewrite (\ref{receivedSignalWithoutHNLoS}) as
\begin{equation}
    \bm{y} = \sqrt{P} \bm{\Lambda}_\phi \bm{\Lambda}_{\bm{\theta}} \mathbf{S} ^ \top  \mathbf{G}^\top \bm{w} + \bm{v},
\end{equation}
with $\mathbf{G} \in \mathbb{C}^{{\color{black} N_{\text{R}}} \times N}$ and $\left[ \mathbf{G} \right]_{r, n} = {\color{black}\sqrt{\rho_{{\text{aR}}}^{n,r}}} {\color{black}\sqrt{\rho_{{\text{Ru}}}^{n,r}}} e^{- j 2 \pi \frac{n B}{N} ({\tau_{{\color{black} \text{a}}r}} + {\color{black} \tau_{r\text{u}}})}$.

Correspondingly, $\bm{\mu}$ can be rewritten as $\bm{\mu} = \sqrt{P} \bm{\Lambda}_\phi \bm{\Lambda}_{\bm{\theta}} \mathbf{S} ^ \top  \mathbf{G}^\top  \bm{\bar{w}}^* $. Furthermore, we can express each term of $\bm{\mu}$ and its conjugate as
\begin{align}
    \bm{\mu}[n] = \bm{\mu}[n]^\top &= \sqrt{P} e^{j (\bm{\theta}[n] + 2 \pi \frac{n}{N} \phi)} [\mathbf{S}^\top]_{n,:} \mathbf{G}^\top  \bm{\bar{w}}^*  \notag \\
    &= \sqrt{P} e^{j (\bm{\theta}[n] + 2 \pi \frac{n}{N} \phi)}  \bm{\bar{w}}^{\rm H} \mathbf{G} [\mathbf{S}]_{:, n},
\end{align}
and 
\begin{equation}
    \bm{\mu}[n]^* = \sqrt{P} e^{-j (\bm{\theta}[n] + 2 \pi \frac{n}{N} \phi)} [\mathbf{S}]_{:, n}^{\rm H} \mathbf{G}^{\rm H}  \bm{\bar{w}}.
\end{equation}
The partial derivatives of $\bm{\mu}[n]$ and $\bm{\mu}[n]^*$ with {regard} to all parameters to be estimated are derived in Appendix \ref{appendixA}. It can be observed that $\bm{\bar{w}}$ is only in the last term of $\frac{\partial \bm{\mu}[n]^*}{\partial \lambda_i}, \forall i$, while $\bm{\bar{w}}^{\rm H}$ appears as the first term of $\frac{\partial \bm{\mu}[n]}{\partial \lambda_j}, \forall j$, for the vector parts. Hence, the product of $\frac{\partial \bm{\mu}[n]^*}{\partial \lambda_i}, \forall i$ and $\frac{\partial \bm{\mu}[n]}{\partial \lambda_j}, \forall j$ would lead to~$\mathbf{W}$.
As a result, as can be seen from (\ref{FIM_summation}), every term of $\bm{\mathrm{FIM}}(\bm{y}; \bm{\lambda})$ is linear to $\mathbf{W}$ because the operations of taking the real parts, summation, and scalar multiplication preserve linearity. Hence, $\mathbf{B}(\bm{p}_{\color{black}s}, \mathbf{W})$ is linear to~$\mathbf{W}$. \qed

\smallskip
\smallskip

According to Proposition 1, (\ref{finalOptimizationProblem}) is a convex SDP problem, and can be {optimally solved} using established convex optimization solvers, such as off-the-shelf CVX toolboxes~\cite{grant2014cvx}. Since {the rank-1 constraint is relaxed via the SDR process,} the obtained matrix $\mathbf{W}$ is likely to have a rank greater than~1. Therefore, decomposing it into the desired RIS phase shift vector $\bm{w}$ is a crucial subsequent problem. As done in \cite{yu2016alternating}, we first perform EVD on the optimal $\mathbf{W}$, then take the largest eigenvector and multiply it with the square root of the largest eigenvalue and normalize the vector, obtaining $\bm{\bar{w}}$. Finally, we calculate the RIS phase shifts by $\bm{w} = \bm{\bar{w}}^*$. The decomposition process can also be accomplished using the Gaussian randomization method, as done in \cite{wu2018intelligent}.


Being an SDP, the problem in (\ref{finalOptimizationProblem}) can be solved using the interior point method in the CVX toolboxes. Thus, the convergence of the proposed RIS configuration is guaranteed. 
The computational complexity of the proposed RIS configuration is {primarily influenced} by the interior point method employed to solve (\ref{finalOptimizationProblem}). The worst-case complexity of the interior point method is {$\mathcal{O}(\max \left\{N_{\color{black} \text{con}}, N_{\color{black} \text{var}}\right\}^4 \sqrt{N_{\color{black} \text{var}}} {\color{black} \log \frac{1}{\epsilon}})$, where $N_{\color{black} \text{con}}$ and $N_{\color{black} \text{var}}$ are the constraint and variable numbers}, respectively, {\color{black}and $\epsilon$ is the convergence accuracy} \cite{luo2010semidefinite}. For the problem in (\ref{finalOptimizationProblem}), ${N_{\color{black} \text{con}}}={\color{black} N_{\text{R}}} + 2$ and ${N_{\color{black} \text{var}} }={\color{black} N_{\text{R}}}^2 + {\color{black}S} (N + 3)^2$. {Suppose that the convergence accuracy is $\epsilon$}. The complexity of solving (\ref{finalOptimizationProblem}) is $\mathcal{O}(({\color{black} N_{\text{R}}}^2 + {\color{black}S}N^2)^{4.5} \log \frac{1}{\epsilon})$. {\color{black}In comparison, the complexity of the state of the art \cite{luan2021phase} is $\mathcal{O}((N_\text{R}^2 + 9 {\color{black}S})^{4.5} \log \frac{1}{\epsilon})$. While incurring a slightly higher complexity, our method improves the PEB by threefold over the benchmark, due to our consideration of CFO and PN, as verified numerically in Section~\ref{section:results}-B.}

\section{Numerical Evaluation}
\label{section:results}
{This section elaborates on the comprehensive {simulation results conducted} to assess the proposed CFO, PN and {\color{black}UE} position estimation and the RIS configuration algorithms.} We {adopt} a 3D Cartesian coordinate system, where the AN is located at ${\color{black} \bm{p}_{\text{a}} } = [\rm 2 \: m, -2 \: m, 1 \: m]^\top$, and the reference point of the RIS is located at $\bm{0}_{3 \times 1} {\rm \: m}$. {The RIS is placed on the $yz$ plane, without loss of generality}. {The} AOI is {set to be} a cube centered at $[\rm 2 \: m, 2 \: m, 0 \: m]^\top$ with the side length of $\rm 1 \: m$. The light speed {is} $c = 3 \times 10^8 \: {\rm m / s}$, and the noise power {is} $\sigma^2 = -109 \; {\rm dBm}$. $N = 32$ subcarriers are used in an OFDM symbol, {the central frequency is ${\color{black} f_{\text{c}}} = 2.8 {\rm \: GHz}$, and the bandwidth is $B = 100 {\rm \: MHz}$.} 
The antenna gains $\color{black} G_{\text{a}}$ and $\color{black}G_{\text{t}}$ are both set to~$1$. Unless otherwise specified, ${\color{black} N_{\text{R}}}=81$ and $\sigma_\Delta^2 = 10^{-3} \; {\rm rad^2}$ are used in the simulations. 

The length of the approximate PN vector $\bm{\eta}$ is $L = 16$. The normalized CFO $\phi$ is uniformly drawn from $[-0.15, 0.15]$, corresponding to a CFO of up to $15 \; {\rm MHz}$. 
The CFO can be caused by frequency mismatches (resulting from severe hardware impairments) between the transmitter (the ANs) and receiver (the UE) when the UE is stationary.\footnote{The CFO can also be a result of the Doppler shift, in which case the normalized CFO is expected to be within a much smaller range, e.g., $[-9.96\times 10^{-5}, 9.96 \times 10^{-5}]$ when the UE moves at the speed of $\rm 120 \; km/h$. 
The proposed localization algorithm can directly apply to this case due to the direct mapping between the Doppler shift and the normalized CFO.} 

We gauge the performance of the proposed algorithms in comparison with the state-of-the-art approaches developed in \cite{luan2021phase}, where a 2D near-field localization problem was solved with the assistance of a RIS. An MLE-based algorithm was designed to {estimate the UE position}, and an SDR-based optimization algorithm was proposed to minimize the CRLB. Note that the CFO and PN were not considered in \cite{luan2021phase}.
The RMSE and the CFO-PN joint MSE are utilized to assess the performance of the localization and joint CFO and PN estimation; see (\ref{RMSE}) and (\ref{jointMSE}).

\subsection{Convergence of Algorithm \ref{alg:A}}
\label{subsec:convergence}

\begin{figure}[t]
    \centering
    \subfloat[Convergence of the GD algorithm for estimating UE position.]{\includegraphics[width=0.9\linewidth]{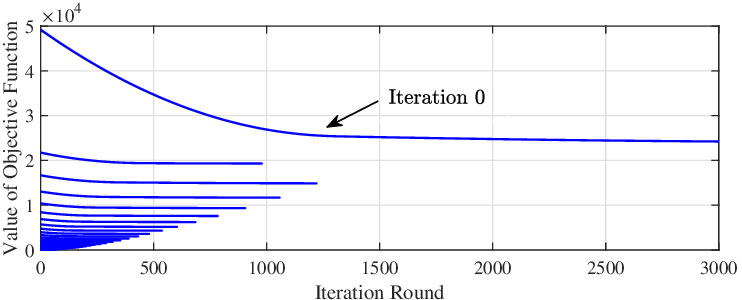}%
    \label{convergence1}}
    \hfil
    \subfloat[Convergence of the overall algorithm.]{\includegraphics[width=0.9\linewidth]{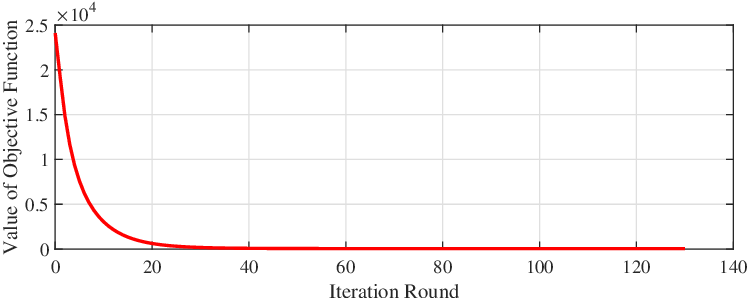}%
    \label{convergence2}}
    \caption{The convergence of the proposed Algorithm 1, where the transmit power is {\color{black}$P=10 {\rm \;  dBm}$}.}
    \label{convergence}
\end{figure}

First, we {\color{black}demonstrate} the convergence of Algorithm \ref{alg:A}. Fig. \ref{convergence} illustrates the {\color{black} case with the transmit power of {\color{black}$-20 {\rm \:  dBm}$},} which corresponds to a receive SNR of approximately {\color{black}$37.5 {\rm \:  dB}$}. Specifically, Fig. \ref{convergence}(a) depicts the change of the objective function when operating the GD algorithm to estimate the UE position. {\color{black}The $k$-th curve ($k=0, 1, \cdots$) from top in Fig. \ref{convergence}(a) shows the convergence of the GD algorithm in the $k$-th iteration of Algorithm \ref{alg:A}. The GD algorithm demonstrates good convergence and effectively reduces the objective function.} Fig. \ref{convergence}(b) focuses on the convergence of the overall algorithm, and different points on the curves represent the {objective function's values} after different rounds of iteration. The GD algorithm generally converges within {\color{black}3,000} iterations {in this setup}, and converges increasingly faster within an iteration. The overall algorithm generally converges within {\color{black}150} iterations. 
Moreover, Table \ref{tab1} shows the average number of iterations of Algorithm \ref{alg:A} {\color{black} and the RMSE of UE localization} under different transmit powers, $N_{\color{black} \text{O} }$ and $N_{\color{black} \text{I} }$ are set to be sufficiently large, and 100 tests are conducted. {\color{black}The results indicate that as the transmit power increases, the number of iterations required decreases while the localization accuracy improves under the proposed algorithm.}
\subsection{Performance of RIS Configuration}

\begin{table}[t]
\color{black}
\begin{center}
\caption{Average number of iterations and UE localization RMSE for Algorithm 1, where $\epsilon_\text{o} = 10^{-7} \times 10^{\rm Round(\log \mathcal{L})}$.}
    \label{tab1}
    \resizebox{0.99\linewidth}{!}{
        \begin{tabular}{| c | c | c | c | c |}
            \hline
            Transmit Power (dBm) & $-20$ & $-10$ & $0$ & $10$ \\
            \hline
            Avg. No. of Iterations & 1238.60 & 327.90 & 195.16 & 137.74\\
            \hline
            RMSE (m)& $0.1590$ & $0.0503$ & $0.0197$ & $0.0090$ \\
            \hline 
        \end{tabular}
    }
\end{center}
\end{table}

\begin{figure*}[t]
\centering
\subfloat[Random RIS phase shift]{\includegraphics[width=0.32\linewidth]{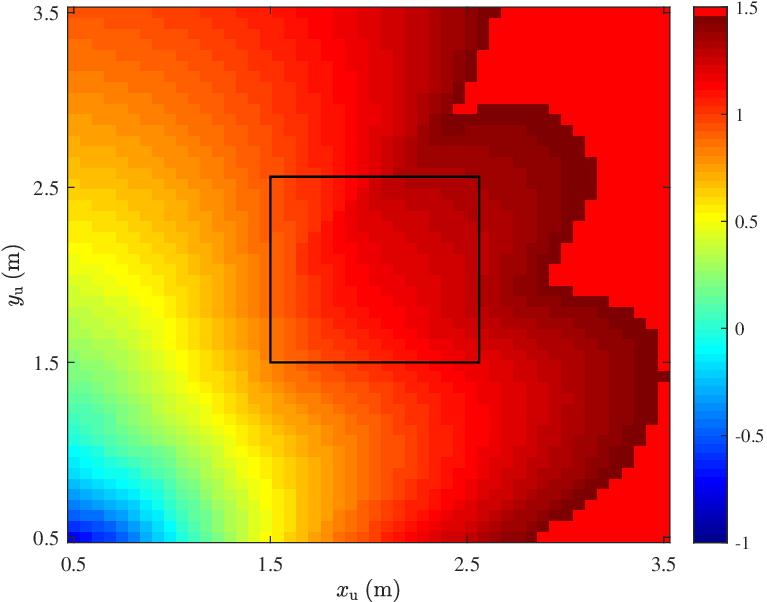}}
\label{PEB_of_AOI_1}
\subfloat[Approach developed in \cite{luan2021phase}]{\includegraphics[width=0.32\linewidth]{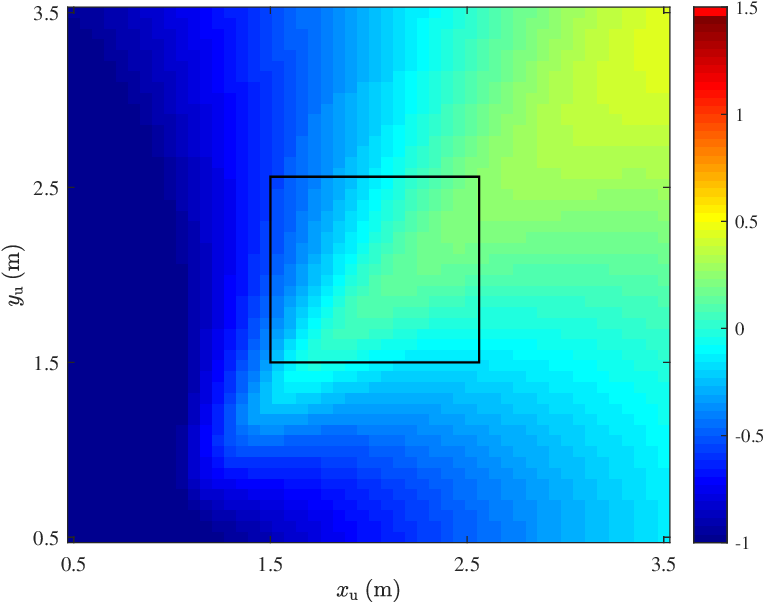}}
\label{PEB_of_AOI_2}
\subfloat[Proposed optimization algorithm]{\includegraphics[width=0.32\linewidth]{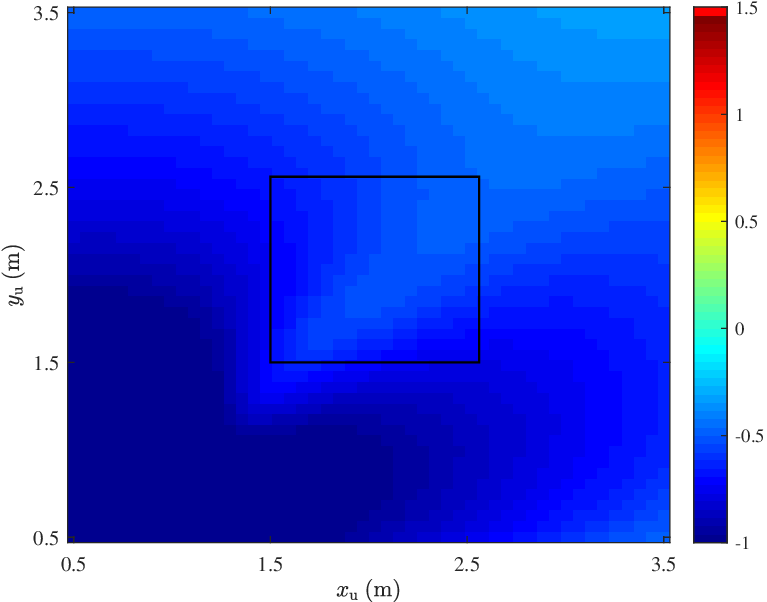}}
\label{PEB_of_AOI_3}
\hfil
\caption{The {\color{black}logarithm of} PEB within AOI {for various RIS configuration strategies}.}
\label{PEB_of_AOI}
\end{figure*}

Next, our attention is directed towards evaluating the performance of the proposed RIS configuration algorithm. Fig.~\ref{PEB_of_AOI} illustrates the PEB near the AOI {across various} RIS phase shift design schemes, where the black squares represent the boundary of AOI. More precisely, we set the transmit power $P$ to $-20 {\rm \: dBm}$, corresponding to a receive SNR of approximately $7.5 {\rm \: dB}$. Without loss of generality, the selected positions in the figures are all located in the $xy$-plane. 

Fig. \ref{PEB_of_AOI} indicates that the proposed optimization algorithm improves the PEB by approximately two orders of magnitude compared to the random RIS phase shifts, and by {\color{black}more than threefold} compared to the situation {\color{black} overlooking} hardware impairments~\cite{luan2021phase}. This demonstrates the {significance} of considering the CFO and PN in the {localization of the UE}, as well as the effectiveness, progressiveness, and superiority of the proposed RIS configuration algorithm. Additionally, {observed from Fig. \ref{PEB_of_AOI}}, the optimized RIS phase shifts not only improve the localization accuracy within the AOI but also significantly improve localization performance in its surrounding areas.


\begin{figure}[t]
\centering
\subfloat[Different RSR volumes]{\includegraphics[width=0.43\linewidth]{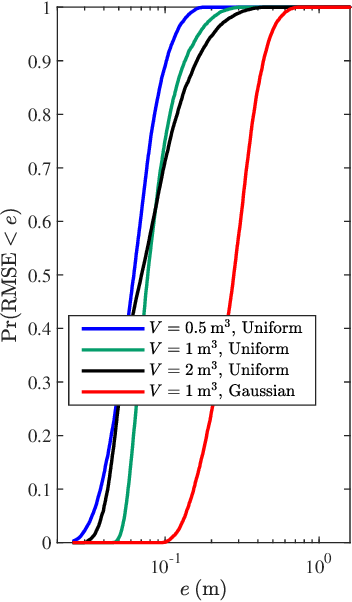}}
\label{cmp_AOI_size}
\subfloat[Different sampling strategies]{\includegraphics[width=0.43\linewidth]{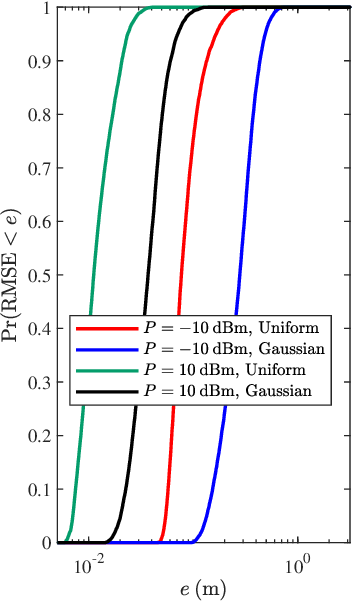}}
\label{cmp_sampling_strategy}
\hfil
\caption{\color{black}The effects of different RSR volumes and random sampling strategies on the CDF of PEB within AOI.}
\label{AOI_and_sampling_strategy}
\end{figure}

{\color{black}
Furthermore, we explore the effects of the size of the \emph{random sampling region} (RSR), and the strategy of random sampling on localization. Note that the RSR with volume $V$, coincides with the AOI by default in the RIS configuration algorithm. Fig. \ref{AOI_and_sampling_strategy}(a) shows the impact of different RSR strategies on the localization performance. By setting the center of the RSR to the center of the AOI, it is observed that as $V$ changes, the CDF of PEB does not vary significantly. This is because the proposed RIS configuration algorithm can improve the localization accuracy within the encompassed region of the RSR, as already revealed in Fig. \ref{PEB_of_AOI}.

Fig. \ref{AOI_and_sampling_strategy}(b) shows the effect of different random sampling strategies on the CDF of PEB within AOI. Uniform and Gaussian sampling are considered. The strategy of random sampling significantly affects the localization performance, and uniform sampling outperforms Gaussian sampling. {\color{black}Moreover, random sampling} affects the performance of the RIS configuration algorithm and the localization accuracy more significantly than the RSR strategy, as also shown in Fig.~\ref{AOI_and_sampling_strategy}(a).}

\subsection{Estimation Accuracy}
\label{EstimationAccuracy}
\begin{figure}[!t]
\centering
\includegraphics[width=0.9\linewidth]{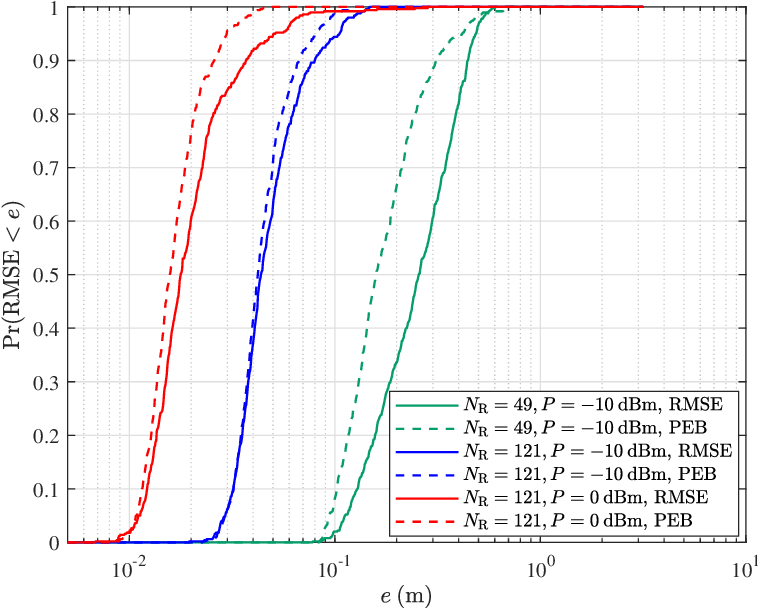}
\caption{The CDF of RMSE and PEB with different ${\color{black} N_{\text{R}}}$ and transmit powers.}
\label{CDF}
\end{figure}

Then, we {concentrate} on the {CFO, PN, and UE position estimation accuracy}. We first illustrate the overall localization accuracy. Fig. \ref{CDF} {illustrates} the \emph{cumulative distribution functions} (CDFs) of RMSE and PEB for 500 random testing positions, under different transmit powers and different numbers of RIS elements. It {is} observed that increasing the RIS size and the transmit power {results in} significant {enhancement of} the overall localization accuracy. Furthermore, when the RIS size is ${\color{black} N_{\text{R}}}=49$, the CDF of the RMSE deviates significantly from the CDF of the PEB, and does not achieve the theoretical localization accuracy. However, when the RIS size is increased to ${\color{black} N_{\text{R}}}=121$, the CDF of the RMSE curve aligns more closely with the CDF of the PEB. This phenomenon can also be observed in Fig. \ref{RMSE_of_different_RIS_size}.

Next, we select a fixed testing position with the coordinates $\rm [1.50 \: m, 2.15 \: m, 0.45 \: m]^\top$ 
to investigate the impact of the transmit power $P$, the RIS size ${\color{black} N_{\text{R}}}$, and the PN variance $\sigma_\Delta^2$ on the {CFO, PN, and UE position estimation accuracy.}

\begin{figure}[!t]
\centering
\includegraphics[width=0.9\linewidth]{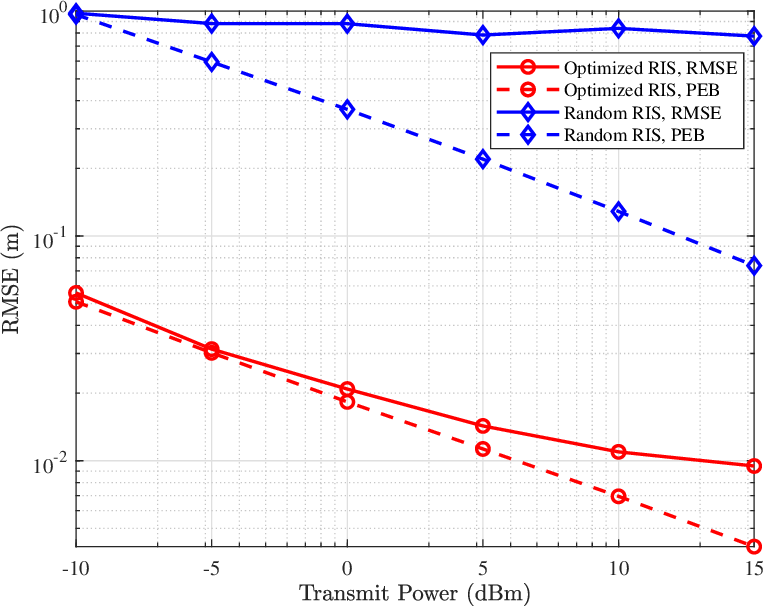}
\caption{The RMSE of $\bm{p}_{\color{black}\text{u}}$ estimation under random and optimized RIS phase shifts.}
\label{RMSE_cmp_with_random_RIS}
\end{figure}

\begin{figure}[!t]
\centering
\includegraphics[width=0.9\linewidth]{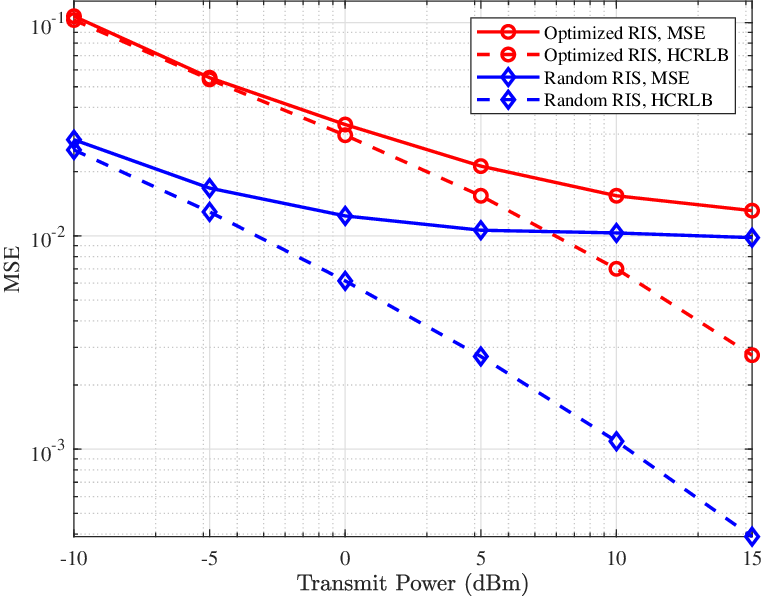}
\caption{${\rm MSE}_{\phi, \bm{\theta}}$ under random and optimized RIS phase shifts.}
\label{PN_CFO_MSE_cmp_with_random_RIS}
\end{figure}

In Figs. \ref{RMSE_cmp_with_random_RIS} and \ref{PN_CFO_MSE_cmp_with_random_RIS}, we assess the {effect} of the optimized RIS {phase shifts} on the {performance of estimation}, compared to {the} random {setup}. From Fig. \ref{RMSE_cmp_with_random_RIS}, the localization performance with random RIS phase shifts does not experience a significant decline as the transmit power increases. At the other extreme, the optimized RIS phase shifts provide an improvement of approximately two orders of magnitude in localization accuracy. {This demonstrates the crucial role of optimizing the RIS phase shifts in localization and underscores the effectiveness of the proposed RIS configuration algorithm.} On the other hand, Fig. \ref{PN_CFO_MSE_cmp_with_random_RIS} shows that {compared to the optimized RIS phase shifts, using random RIS phase shifts yields even lower joint MSE of CFO and PN, although the curve does not experience any significant decline as the transmit power increases in this case.} This occurs because we extract the PEB part of HCRLB and set it as the RIS optimization objective. Thus, the estimation performance of the CFO and PN parameters is ignored.

\begin{figure}[!t]
\centering
\includegraphics[width=0.9\linewidth]{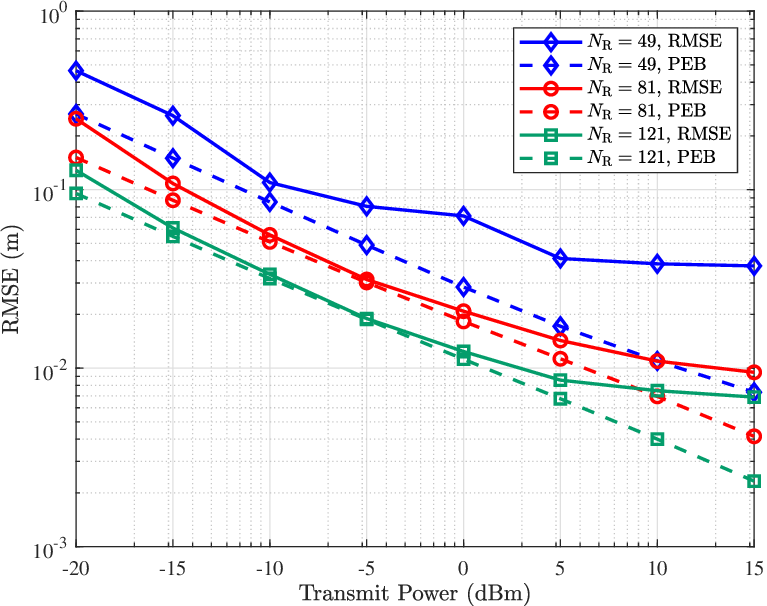}
\caption{The RMSE of $\bm{p}_{\color{black}\text{u}}$ estimation at ${\color{black} N_{\text{R}}}=49$, ${\color{black} N_{\text{R}}}=81$, and ${\color{black} N_{\text{R}}}=121$.}
\label{RMSE_of_different_RIS_size}
\end{figure}

\begin{figure}[!t]
\centering
\includegraphics[width=0.9\linewidth]{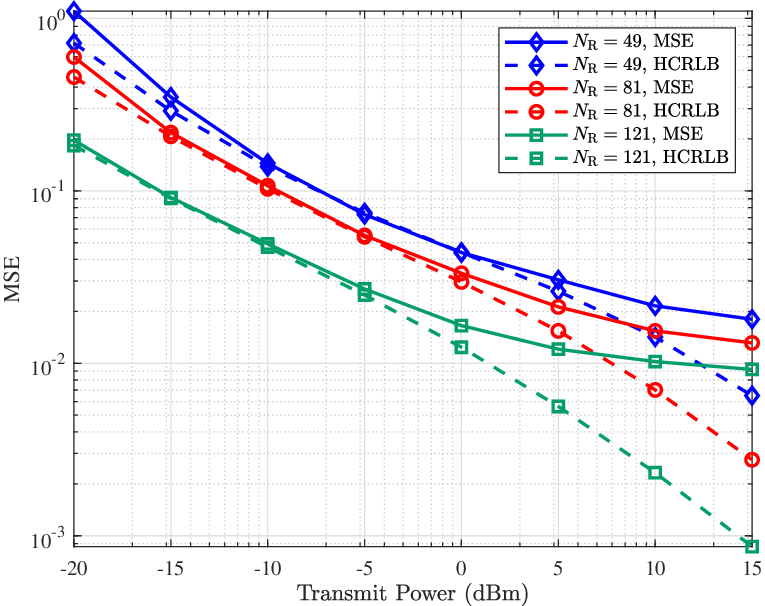}
\caption{${\rm MSE}_{\phi, \bm{\theta}}$ at ${\color{black} N_{\text{R}}}=49$, ${\color{black} N_{\text{R}}}=81$, and ${\color{black} N_{\text{R}}}=121$.}
\label{PN_CFO_MSE_of_different_RIS_size}
\end{figure}

Figs. \ref{RMSE_of_different_RIS_size} and \ref{PN_CFO_MSE_of_different_RIS_size} depict the curves of the RMSE for UE localization and MSE for the estimation of $\underline{\bm{\gamma}}$, respectively, under different transmit powers, where {${\color{black} N_{\text{R}}}$} is {\color{black}set to} 49, 81, or 121. It is observed that increasing the transmit power and/or the {RIS element number} {results} in gains in all parameters to be estimated. However, as the transmit SNR increases, the RMSE and MSE do not decrease as sharply as HCRLB{\color{black},} because of the approximate PN vector, as well as the Taylor series approximation considered when deriving closed-form expressions for the CFO and PN.

\begin{figure}[!t]
\centering
\includegraphics[width=0.9\linewidth]{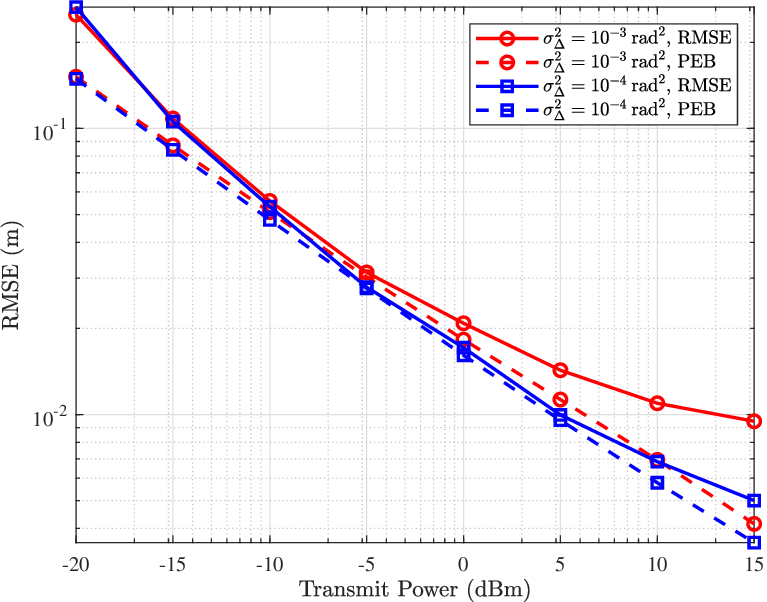}
\caption{${\rm RMSE}_\text{u}$ at $\sigma_\Delta^2=10^{-3} {\rm \: rad^2}$ and $\sigma_\Delta^2=10^{-4} {\rm \: rad^2}$.}
\label{RMSE_of_different_sigma_Delta_2}
\end{figure}

\begin{figure}[!t]
\centering
\includegraphics[width=0.9\linewidth]{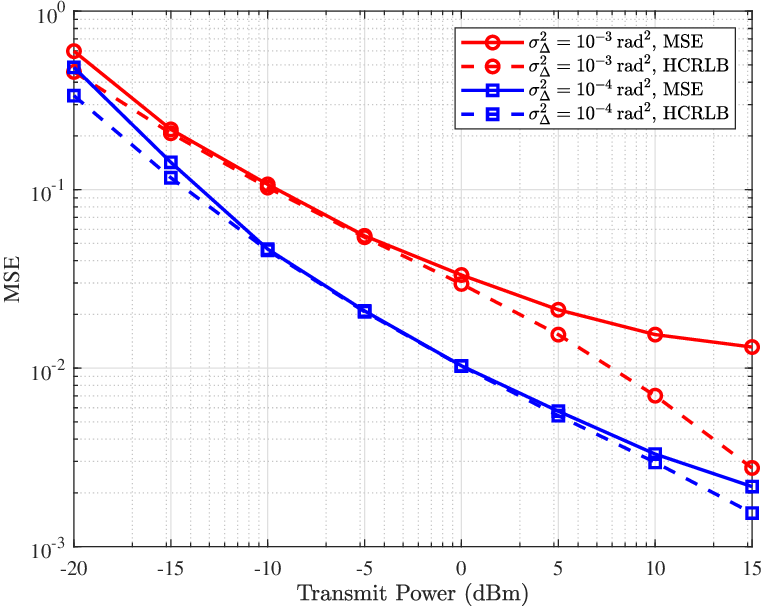}
\caption{${\rm MSE}_{\phi, \bm{\theta}}$ at $\sigma_\Delta^2=10^{-3} {\rm \: rad^2}$ and $\sigma_\Delta^2=10^{-4} {\rm \: rad^2}$.}
\label{PN_CFO_MSE_of_different_sigma_Delta_2}
\end{figure}

Figs. \ref{RMSE_of_different_sigma_Delta_2} and \ref{PN_CFO_MSE_of_different_sigma_Delta_2} plot the RMSE for localization and MSE for estimation of $\underline{\bm{\gamma}}$, respectively, when ${\color{black} N_{\text{R}}}$ is set to 81, where the PN variance is $\sigma_\Delta^2=10^{-3} {\rm \: rad^2}$ or $\sigma_\Delta^2=10^{-4} {\rm \: rad^2}$ {\color{black}\cite{mehrpouyan2012joint, ding2024parameter, gavert2022estimation, du2024maximum}. $\sigma_\Delta^2 = 10^{-3} \: {\rm rad^2}$ is a considerably high PN variance \cite{mehrpouyan2012joint}.} The results demonstrate that decreasing the PN covariance leads to significant improvements in the estimation performance of the CFO, PN, and UE position. Another important finding is that a smaller PN variance would make the RMSE and MSE curves more closely aligned with the HCRLB curves because of the more accurately approximated parameters when updating the CFO and PN parameters.

\begin{figure}[!t]
\centering
\includegraphics[width=0.9\linewidth]{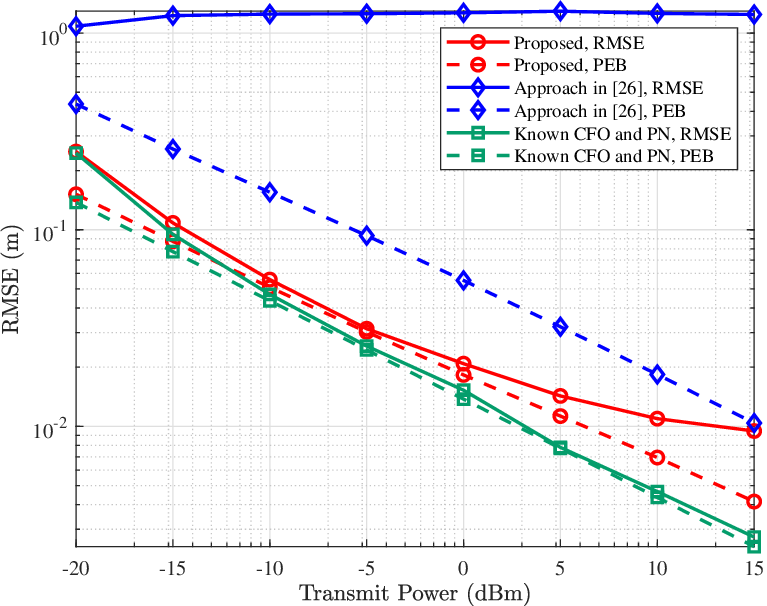}
\caption{Benchmarking the performance of the proposed algorithms against the state-of-the-art approach in \cite{luan2021phase}, {\color{black} as well as the case with perfectly known CFO and PN}.}
\label{RMSE_cmp_with_TVT}
\end{figure}

{\color{black}Furthermore,} we compare the proposed overall algorithm of the CFO, PN, and UE position estimation and the RIS configuration, with the benchmark algorithm developed in \cite{luan2021phase}. It is shown in Fig. \ref{RMSE_cmp_with_TVT} that, after optimization, the theoretical localization accuracy of our proposed algorithm outperforms the benchmark, which confirms the conclusion drawn in Fig. \ref{PEB_of_AOI}(b). As for the localization RMSE, the benchmark shows a large gap with its corresponding PEB, and its performance shows no demonstrable correlation with the transmit power because the estimation algorithm totally overlooked the CFO and PN. In comparison, the proposed algorithm significantly improves the localization accuracy by approximately one order of magnitude under relatively low transmit powers of below $-10 {\rm \: dBm}$, and by two orders of magnitude under high transmit powers of over $10 {\rm \: dBm}$. 

{\color{black} Fig. \ref{RMSE_cmp_with_TVT} also depicts the localization accuracy when the CFO and PN values are perfectly known. Compared to the proposed algorithm and the benchmark, effective mitigation of the CFO and PN effects can play a crucial role in improving localization accuracy. Furthermore, the utilization of our algorithm results in localization accuracy that is very close to the optimal value with known CFO and PN.}

{\color{black}

\begin{figure}[!t]
\centering
\includegraphics[width=0.9\linewidth]{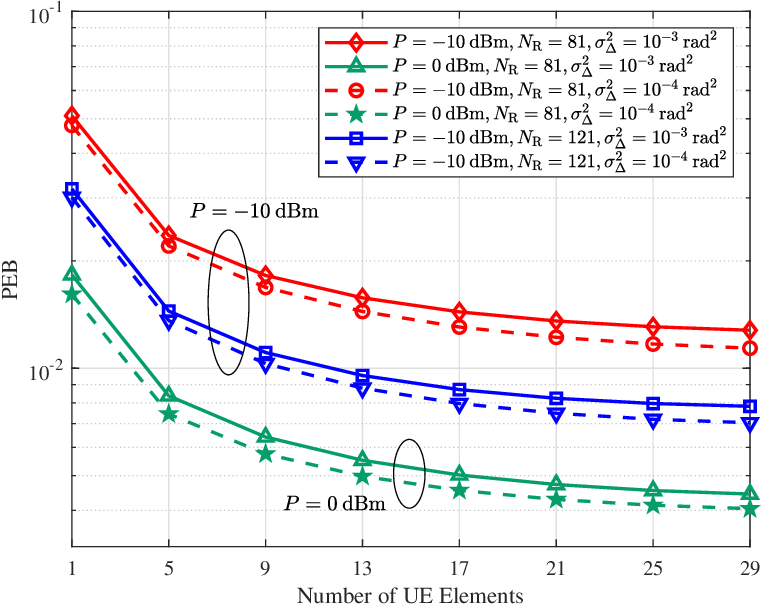}
\caption{\color{black}The impact of the number of antenna elements at the UE on the PEB under different settings of $P$ and $\sigma_\Delta^2$.}
\label{SIMO_PEB}
\end{figure}

\begin{figure}[!t]
\centering
\includegraphics[width=0.9\linewidth]{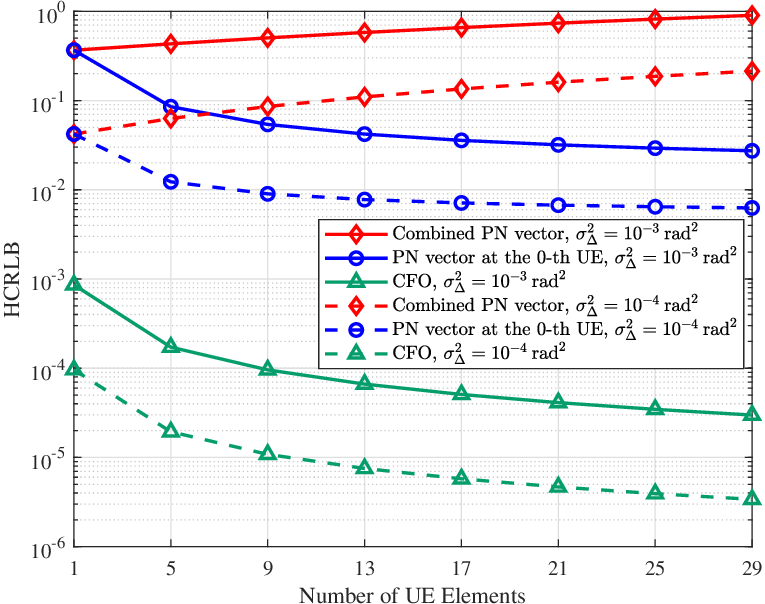}
\caption{\color{black}The impact of the number of antenna elements at the UE on the HCRLB of CFO and PN under different settings of $\sigma_\Delta^2$.}
\label{SIMO_HCRLB}
\end{figure}
\color{black}

{\color{black}In Figs. \ref{SIMO_PEB} and \ref{SIMO_HCRLB}, we investigate} the impact of the number of antennas at the UE, $M$, on UE localization accuracy. The antennas are arranged in a uniform linear array along the $y$-axis, with spacing of $d$. Combining signals received by all antennas, we have $\bm{y}_\text{com} = \bm{\mu}_\text{com} + \bm{v}_\text{com}$, where $\bm{y}_\text{com} = \left[\bm{y}_0^\top, \bm{y}_1^\top, \cdots, \bm{y}_{M-1}^\top \right]^\top$, $\bm{\mu}_\text{com} = \left[\bm{\mu}_0^\top, \bm{\mu}_1^\top, \cdots, \bm{\mu}_{M-1}^\top \right]^\top$, and $\bm{v}_\text{com} = \left[\bm{v}_0^\top, \bm{v}_1^\top, \cdots, \bm{v}_{M-1}^\top \right]^\top$. $\bm{y}_m$, $\bm{\mu}_m$, and $\bm{v}_m$, $m = 0, 1, \cdots, M-1$, represent the received signal, noise-free received signal, and combined noise at the $m$-th antenna of the UE, respectively. $\bm{\mu}_m$ can be further expressed as $\bm{\mu}_m = \sqrt{P} \bm{\Lambda}_\phi \bm{\Lambda}_{\bm{\theta}_m} \mathbf{S} ^ \top \bm{h}_m$, where $\bm{h}_m$ can be obtained by \color{black}substituting the coordinates of the UE's $m$-th antenna into~(\ref{h_los}). $\bm{\theta}_m,\, m = 0, 1, \cdots, M - 1$, is the \color{black}PN vector at the UE's $m$-th antenna and is independent and identically distributed~\cite{6212402}.

Fig. \ref{SIMO_PEB} plots the PEB with the increase in $M$. The impacts of the PN variance $\sigma_\Delta^2$ and the RIS size $N_\text{R}$ are considered. It is observed that as $M$ increases, the analytical localization accuracy error bound, i.e., PEB, decreases significantly. This indicates that increasing the number of UE antennas, like increasing the transmit power and the number of RIS elements, is effective in improving localization accuracy. Under our experimental settings, compared to a single-antenna UE, adding 4 receive antennas yields a greater localization performance gain than adding 40 RIS elements; adding 8 receive antennas yields a localization gain greater than increasing the transmit power by 10 dBm.

\color{black}
Fig. \ref{SIMO_HCRLB} depicts the respective HCRLBs of the combined PN vector, $\bm{\theta}_{\rm com} \triangleq \left[\bm{\theta}_0, \bm{\theta}_1, \cdots, \bm{\theta}_{M-1}\right]$, the PN vector at the $0$-th UE, $\bm{\theta}_0$, and the CFO, $\phi$, under different $\sigma_\Delta^2$ settings. It is observed that the HCRLBs of $\bm{\theta}_{0}$ and $\phi$ decrease as $M$ increases, due to the additional information introduced by increasing the antennas at the UE. The HCRLB of $\phi$ decreases faster than that of $\bm{\theta}_0$, because the rise of $M$ increases observations of $\phi$ but does not increase observations of $\bm{\theta}_0$. Nevertheless, the improved estimation accuracy of $\phi$ and UE position indirectly improves the estimation accuracy of $\bm{\theta}_0$. By contrast, the HCRLB of the combined PN vector increases with $M$, as the increase of $M$ results in more PN vectors that require estimation. Even though the estimation of the PN on each individual antenna of the UE becomes more accurate, considering all PN vectors, the overall estimation accuracy deteriorates.}

\subsection{\color{black}IEEE 802.11n Setup}
\begin{figure}[!t]
\centering
\subfloat[\color{black}Random RIS phase shift]{\includegraphics[width=0.9\linewidth]{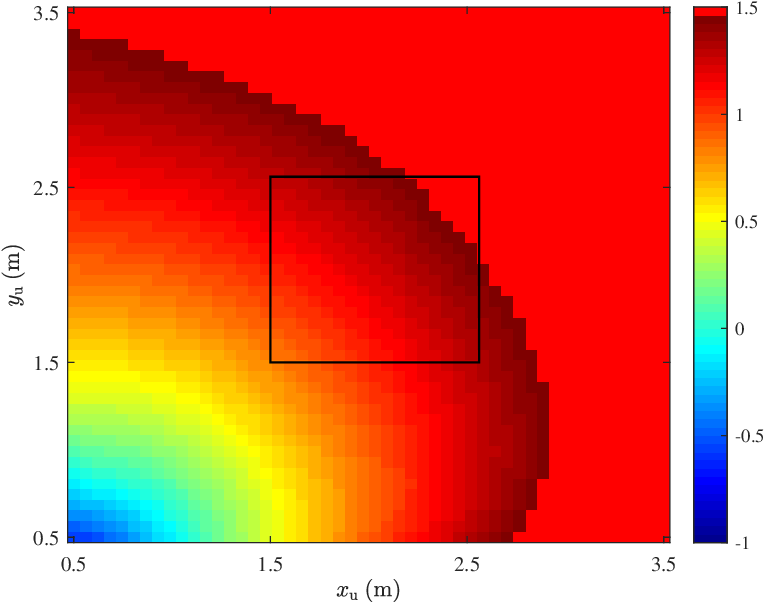}}
\label{WiFi_PEB_of_AOI_1}
\hfill
\subfloat[\color{black}Proposed optimization algorithm]{\includegraphics[width=0.9\linewidth]{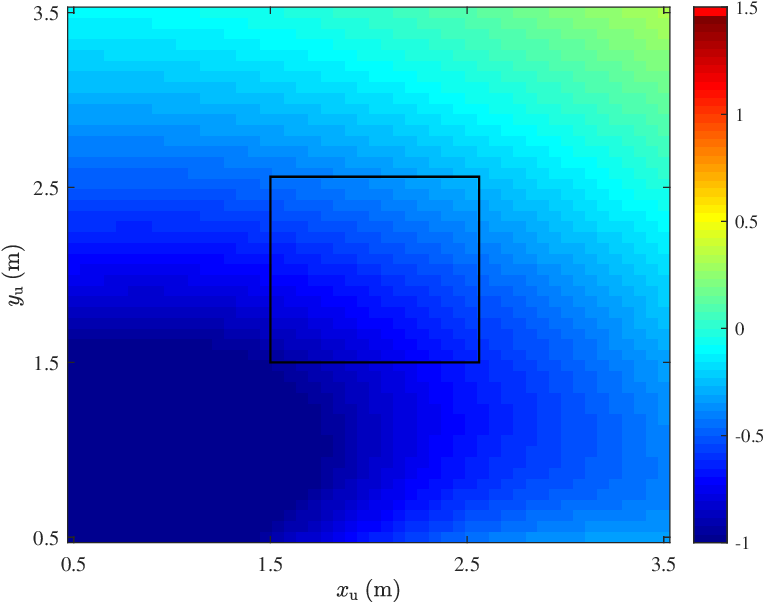}}
\label{WiFi_PEB_of_AOI_2}
\caption{\color{black}The logarithm of PEB within AOI for various RIS configuration strategies under standard IEEE 802.11n settings.}
\label{WiFi_PEB_of_AOI}
\end{figure}

\begin{figure}[!t]
\centering
\includegraphics[width=0.9\linewidth]{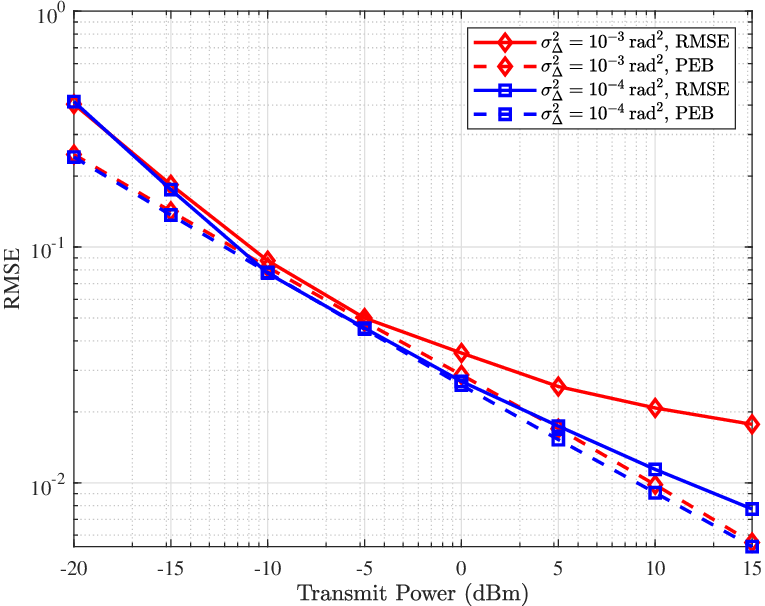}
\caption{\color{black}${\rm RMSE}_\text{u}$ at $\sigma_\Delta^2=10^{-3} {\rm \: rad^2}$ and $\sigma_\Delta^2=10^{-4} {\rm \: rad^2}$ under standard IEEE 802.11n settings.}
\label{WiFi_RMSE_of_different_sigma_Delta_2}
\end{figure}

\begin{figure}[!t]
\centering
\includegraphics[width=0.9\linewidth]{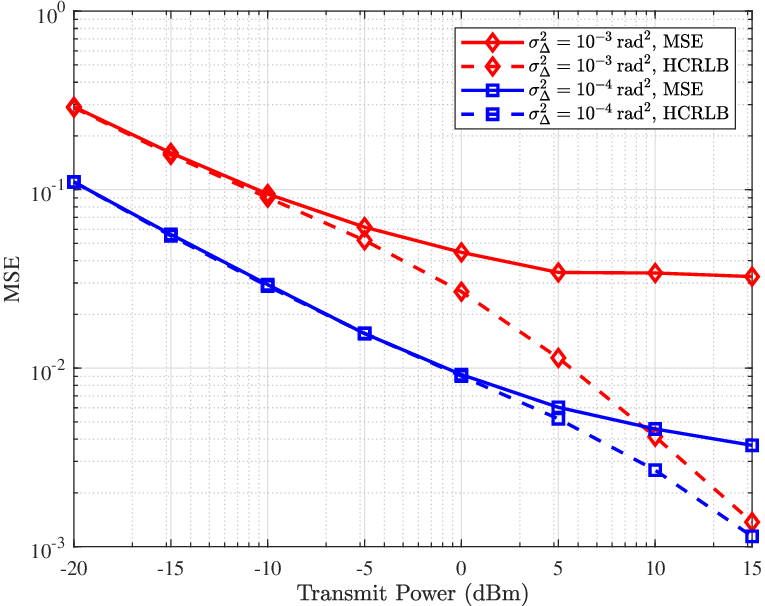}
\caption{\color{black}${\rm MSE}_{\phi, \bm{\theta}}$ at $\sigma_\Delta^2=10^{-3} {\rm \: rad^2}$ and $\sigma_\Delta^2=10^{-4} {\rm \: rad^2}$ under standard IEEE 802.11n settings.}
\label{WiFi_PN_CFO_MSE_of_different_sigma_Delta_2}
\end{figure}

{\color{black}
\color{black}
Following the IEEE 802.11n (Wi-Fi 4) \cite{5307322} settings, the carrier frequency $f_{\color{black}\text{c}}$ is set to $2.4 \: {\rm GHz}$, $N=64$ subcarriers with $8$ null subcarriers are utilized, and the bandwidth $B$ \color{black}is \color{black} $20 \: {\rm MHz}$. Fig. \ref{WiFi_PEB_of_AOI} plots the logarithm of PEB near the AOI under both random and optimized RIS configurations. The proposed algorithm improves the PEB by approximately two \color{black}orders of magnitude, compared to the random RIS phase shifts. Figs. \ref{WiFi_RMSE_of_different_sigma_Delta_2} and \ref{WiFi_PN_CFO_MSE_of_different_sigma_Delta_2} plot the RMSE and joint MSE of CFO and PN under different $\sigma_\Delta^2$ settings, respectively. It is observed that decreasing the PN covariance improves the estimation performance of the CFO, PN, and UE position significantly. Overall, Figs. \ref{WiFi_PEB_of_AOI}, \ref{WiFi_RMSE_of_different_sigma_Delta_2}, and \ref{WiFi_PN_CFO_MSE_of_different_sigma_Delta_2} provide consistent conclusions with those drawn in Figs. \ref{PEB_of_AOI}, \ref{RMSE_of_different_sigma_Delta_2}, and \ref{PN_CFO_MSE_of_different_sigma_Delta_2}, demonstrating the ability of our algorithms to be extended to practical systems.}
\section{Conclusion}
\label{section:conclusion}

This paper investigates the optimization of RIS phase shifts and the joint estimation of CFO, PN, and UE position under imperfect hardware conditions for an OFDM system in a near-field  localization scenario. An AO-based, iterative joint estimation algorithm based on the MAP criterion and GD algorithm has been proposed.
Extensive simulations have demonstrated that by optimizing the RIS phase shifts, the theoretical localization accuracy has been improved by two orders of magnitude. Moreover, the achievable localization accuracy can be lower than $\rm 10^{-2} \: m$. These findings corroborate the {effectiveness} of the proposed algorithms.
To further improve the estimation performance and support a larger AOI, the optimization algorithm used in this paper requires averaging a large number of PEBs related to random positions with high computational complexity. Hence, it becomes imperative to utilize an optimization algorithm with reduced complexity in this scenario. This will be our future work.

\begin{appendices}
        
\section{Derivation of partial derivatives of $\bm{\mu}[n]$ and $\bm{\mu}[n]^*$}
\label{appendixA}

The partial derivatives of $\bm{\mu}[n]$ with {regard} to all parameters to be estimated are derived as follows
\begin{subequations} \label{eq:mu_n_partials}
\begin{align}
\frac{\partial \bm{\mu}[n]}{\partial \phi} &= \sqrt{P} j \frac{2 \pi n}{N}e^{j (\bm{\theta}[n] + 2 \pi \frac{n}{N} \phi)} \bm{\bar{w}}^{\rm H} \mathbf{G} [\mathbf{S}]_{:, n}, \\
 \frac{\partial \bm{\mu}[n]}{\partial \bm{\theta}[n]} &= \sqrt{P} j e^{j (\bm{\theta}[n] + 2 \pi \frac{n}{N} \phi)} \bm{\bar{w}}^{\rm H} \mathbf{G} [\mathbf{S}]_{:, n}, \\
\frac{\partial \bm{\mu}[n]}{\partial \bm{\theta}[m]} &= 0, m \neq n, \\
\frac{\partial \bm{\mu}[n]}{\partial {\color{black}d_\text{R}}} &= \sqrt{P} e^{j (\bm{\theta}[n] + 2 \pi \frac{n}{N} \phi)}  \bm{\bar{w}}^{\rm H}  
 \frac{\partial \mathbf{G}}{\partial {\color{black}d_\text{R}}} [\mathbf{S}]_{:, n}, \\
\frac{\partial \bm{\mu}[n]}{\partial {\color{black}\phi_\text{az}}} &=  \sqrt{P} e^{j (\bm{\theta}[n] + 2 \pi \frac{n}{N} \phi)}  \bm{\bar{w}}^{\rm H} \frac{\partial \mathbf{G}}{\partial {\color{black}\phi_\text{az}}} [\mathbf{S}]_{:, n}, \\
\frac{\partial \bm{\mu}[n]}{\partial {\color{black}\phi_\text{el}}} &= \sqrt{P} e^{j (\bm{\theta}[n] + 2 \pi \frac{n}{N} \phi)}  \bm{\bar{w}}^{\rm H} \frac{\partial \mathbf{G}}{\partial {\color{black}\phi_\text{el}}} [\mathbf{S}]_{:, n},
\end{align}
\end{subequations}
and the partial derivatives of $\bm{\mu}[n]^*$ with {regard} to all parameters to be estimated are derived as follows
\begin{subequations} \label{eq:mu_n_conj_partials}
\begin{align}
\frac{\partial \bm{\mu}[n]^*}{\partial \phi} &= - j \sqrt{P} \frac{2 \pi n}{N}e^{-j (\bm{\theta}[n] + 2 \pi \frac{n}{N} \phi)}  [\mathbf{S}]_{:, n}^{\rm H} \mathbf{G}^{\rm H}  \bm{\bar{w}}, \\
\frac{\partial \bm{\mu}[n]^*}{\partial \bm{\theta}[n]} &= - j \sqrt{P} e^{-j (\bm{\theta}[n] + 2 \pi \frac{n}{N} \phi)}  [\mathbf{S}]_{:, n}^{\rm H} \mathbf{G}^{\rm H}  \bm{\bar{w}}, \\
\frac{\partial \bm{\mu}[n]^*}{\partial \bm{\theta}[m]} &= 0, m \neq n, \\
\frac{\partial \bm{\mu}[n]^*}{\partial {\color{black}d_\text{R}}} &=  \sqrt{P} e^{-j (\bm{\theta}[n] + 2 \pi \frac{n}{N} \phi)} [\mathbf{S}]_{:, n}^{\rm H} \left( \frac{\partial \mathbf{G}}{\partial {\color{black}d_\text{R}}}\right)^{\rm H}  \bm{\bar{w}}, \\
\frac{\partial \bm{\mu}[n]^*}{\partial {\color{black}\phi_\text{az}}} &=  \sqrt{P} e^{-j (\bm{\theta}[n] + 2 \pi \frac{n}{N} \phi)} [\mathbf{S}]_{:, n}^{\rm H} \left( \frac{\partial \mathbf{G}}{\partial {\color{black}\phi_\text{az}}}\right)^{\rm H}  \bm{\bar{w}}, \\
\frac{\partial \bm{\mu}[n]^*}{\partial {\color{black}\phi_\text{el}}} &=  \sqrt{P} e^{-j (\bm{\theta}[n] + 2 \pi \frac{n}{N} \phi)} [\mathbf{S}]_{:, n}^{\rm H} \left( \frac{\partial \mathbf{G}}{\partial {\color{black}\phi_\text{el}}}\right)^{\rm H}  \bm{\bar{w}}.
\end{align}
\end{subequations}
\end{appendices}

\bibliographystyle{IEEEtran}
{\bibliography{IEEE_REF}}


\end{document}